\documentclass[12pt]{article}
\usepackage[margin=1in]{geometry}
\usepackage{amsmath,amsthm,amssymb}
\usepackage{setspace}
\usepackage{natbib}
\usepackage{bm}
\usepackage{mathrsfs}
\usepackage{comment}
\usepackage{booktabs}
\usepackage{multirow}
\usepackage{enumitem}

\usepackage{tikz}
\usetikzlibrary{arrows.meta,positioning}

\newcommand{\Data}{\mathcal D}

\newcommand{\E}{\mathbb{E}}
\newcommand{\Exp}{\operatorname{Exp}}

\newcommand{\Identity}{\mathrm{I}}

\newcommand{\Normal}{\operatorname{Normal}}

\newcommand{\sX}{\mathcal X}

\newcommand{\Tree}{\mathcal T}

\newcommand{\Var}{\operatorname{Var}}
\newcommand{\R}{\mathbb R}

\newcommand{\tony}[1]{#1}

\newenvironment{added}{}{}

\newtheorem{theorem}{Theorem}

\theoremstyle{definition}
\newtheorem{remark}{Remark}

\usepackage{xcolor}  
\usepackage{hyperref}

\hypersetup{
    colorlinks=true,            
    linkcolor=teal,             
    citecolor=blue!70,    
    filecolor=magenta,          
    urlcolor=cyan               
}

\newenvironment{tocheck}{\color{red}}{}
\newenvironment{new}{}{}

\begin{document}

\title{Decision Theoretic Subgroup Detection With Bayesian Machine Learning}
\date{}

\author{
  Entejar Alam\thanks{\texttt{entejar@utexas.edu}, Equal contribution},
  Poorbita Kundu\thanks{\texttt{poorbitakundu@gmail.com}, Equal contribution},
  and Antonio R. Linero\thanks{\texttt{antonio.linero@austin.utexas.edu}}
}

\maketitle

\begin{abstract}
  We consider the problem of identifying promising subpopulations in terms of treatment effectiveness or treatment effect heterogeneity, from a Bayesian decision theoretic perspective. We first show that a straightforward application of Bayesian decision theory to subgroup detection leads to counterintuitive risk-seeking (RS) behavior. Motivated by this observation, we introduce the \emph{Bayesian Risk-Aware Inference and Detection of Subgroups} (BRAIDS) utility and use it to perform subgroup selection and post-selection inference. The BRAIDS utility interpolates between risk-seeking (RS) and risk-averse (RA) identification of subgroups, with a variant of the virtual twins algorithm as its risk-neutral midpoint. We also argue that effective subgroup estimation and inference require the use of \emph{regularization priors} to safeguard inferences from the winner's curse. We provide empirical evidence that posterior credible intervals for subgroup effects can still attain nominal coverage levels, provided that an appropriate prior distribution is chosen. The proposed framework is illustrated on data from a clinical trial assessing the efficacy of canagliflozin as a treatment for type 2 diabetes. 

  \vspace{1em}

  
\end{abstract}

\doublespacing

\section{Introduction}\label{introduction}

In recent years, estimating heterogeneous causal effects in settings where different individuals respond differently to the same treatment has become an important problem for guiding decision making across many domains, from healthcare and education to economics and public policy \citep{imai2011estimation, hitsch2024heterogeneous, yeager2019national}. The Bayesian framework provides a coherent and flexible way to model treatment effect heterogeneity, allowing researchers to incorporate prior knowledge and apply principled regularization through structured priors \citep{hill2011bayesian, hahn2018regularization, hahn2020bayesian, shin2024treatment, linero2023and}. Bayesian tree-based models, such as BART and its extensions, have shown strong empirical performance in capturing heterogeneity while producing calibrated posterior uncertainty \citep{chipman2010bart, dorie2019automated, hahn2020bayesian, woody2021model}. Additionally, a growing body of work has leveraged machine learning methods to detect and analyze treatment effect heterogeneity, with theoretical guarantees even when using ``black box'' algorithms \citep{athey2016recursive, kennedy2023towards, nie2021quasi}.

Despite these advances, precisely estimating heterogeneous treatment effects remains highly challenging \citep{thal2023causal}, especially in high-dimensional or nonparametric settings with limited data and noisy outcomes. In our experience, while appropriately designed machine learning methods can be very useful for estimating \emph{average} treatment effects, we have found that treatment effect heterogeneity is often very sensitive to the degree of regularization used, and that it is difficult to estimate the degree of heterogeneity at the individual level. For example, Figure~\ref{fig:plot_regret_ptau} displays estimated treatment effects on the same dataset using different methods, which are subsequently used to construct plausible data generating mechanisms for a simulation study; we see that different, reasonable, methods estimate very different amounts of treatment effect heterogeneity.

It may therefore be more practical to instead focus on the objective of identifying meaningful and \emph{coarse} subgroups with different treatment responses. In addition to being easier, this goal often aligns better with the needs of decisionmakers, who typically require inferences that are interpretable and actionable. This strategy has proven particularly successful in clinical trials \citep{foster2011subgroup, jones2011bayesian, nugent2019bayesian} and policy evaluation \citep{kitagawa2018who, athey2021policy}. Subgroup effect estimation is a middle ground between population average causal effect estimation on the one hand and conditional average causal effect estimation on the other.

This paper develops a framework for subgroup identification
using Bayesian machine learning and decision theory. We formalize the problem using Bayesian decision theory, and identify promising subgroups
by maximizing an associated posterior expected utility function. After identifying these subgroups, we perform inference on the average treatment effect within each subgroup. We make the following contributions:

\begin{enumerate}[leftmargin=1em]

\item We show that, when Bayesian decision theory is used, the most obvious choice of utility function leads to counterintuitive \emph{risk seeking} (RS) behavior, tending to prefer subgroups for which there is a large amount of uncertainty in the estimated treatment effects. We then introduce the BRAIDS (Bayesian Risk-Aware Inference and Detection of Subgroups) utility,  which embeds the standard utility function within a larger class of ``multi-stage'' utilities that also allow for \emph{risk averse} (RA) and \emph{risk neutral} (RN) behaviors.


  \item We show that regularization via hierarchical priors is critical for Bayesian post-selection inference, and we show how to regularize Bayesian linear regression and Bayesian causal forest \citep{hahn2020bayesian} models. By contrast, when flat priors are used, we see very poor Frequentist coverage of credible intervals.
    We find empirically that credible intervals from appropriately regularized models perform well by Frequentist measures even when the subgroups are estimated from the data.
    This is a large benefit, as it is more efficient than data splitting \citep[see][for a review of post-selection inference]{kuchibhotla2022post}.

  \item We show in Theorem~\ref{thm:bart-hetero} that certain Bayesian causal forest models have the attractive property of inducing priors on the degree of treatment effect heterogeneity whose mean is \emph{invariant} to the distribution of the covariates. Hence, adding or transforming covariates does not greatly affect our prior beliefs about treatment effect heterogeneity, regardless of their correlation structure. This property is \emph{not} shared by Bayesian linear models.

\end{enumerate}

\subsection{Notation and The Canagliflozin Trial}
\label{notation}

We will describe the setting in terms of a clinical trial that investigated the efficacy of canagliflozin as a treatment for type 2 diabetes mellitus (T2DM). Canagliflozin has been shown to reduce the risk of cardiovascular and renal events in patients with T2DM; results from preliminary trials, however, find a heterogeneous treatment effect across different subpopulations of patients, and it is of interest to identify the subgroups of patients that respond differently to treatment.
The trial protocols included prespecified subgroup analyses that were to be performed, giving a natural point of comparison for estimated subgroups. 

We operate within the Rubin causal model \citep{rubin1974estimating} with potentially observed data $\{Y_i(0), Y_i(1), A_i, X_i : i = 1,\ldots,N\}$ and observed data $\mathcal D = \{Y_i, A_i, X_i : i = 1,\ldots,N\}$. In the canagliflozin trial, $Y_i$ is the change from baseline in glycated hemoglobin (HBA$_{1c}$), $A_i = 1$ or $A_i = 0$ according to whether an individual was assigned to a particular dosage of canagliflozin or to placebo, and $X_i$ is a collection of pretreatment effect modifiers of interest: age, race, baseline HBA$_{1c}$, sex, and ethnicity. We make the following causal assumptions: (i) consistency and the stable unit treatment value assumption that $Y_i = Y_i(A_i)$; (ii) strong ignorability of the assignment mechanism $[\{Y_i(0), Y_i(1)\} \perp A_i \mid X_i]$, i.e., the potential outcomes are independent of the assigned treatment given $X_i$; and (iii) positivity of the treatment assignment, with $\delta \le \Pr(A_i = a \mid X_i = x) \le 1 - \delta$ for some positive $\delta$. Because treatment was randomized, we know assumptions (ii) and (iii) hold for the canagliflozin trial.

\tony{The covariates $X_i = (X_{i1}, \ldots, X_{iP})^\top$ are assumed to be sampled from some distribution $F_X$ on $\sX \subseteq \R^P$ independently for $i = 1,\ldots,N$.} We also define the probability distribution of \([X_i \mid X_i \in G]\) as \(F_{X \mid G}(dx)\). We let \(e(x) = \Pr(A_i = 1 \mid X_i = x)\) denote the \emph{propensity score} that determines the probability of observational unit \(i\) receiving treatment $a = 1$. While this work focuses primarily on randomized clinical trials, our methodology is also applicable to observational studies, in which case $e(x)$ is not known. We let \(\tau(x) = \mathbb E\{Y_i(1) - Y_i(0) \mid X_i = x\}\) denote the treatment effect function and we let $ \tau(G) = \mathbb E\{Y_i(1) - Y_i(0) \mid X_i \in G\} = \frac{ \sum_{i: X_i \in G} \tau(X_i)}{\sum_{i: X_i \in G} 1}$ denote the in-sample subgroup treatment effect for subset $G$.

\subsection{Related Work}

Our work builds on a large literature on heterogeneous treatment effect estimation and subgroup identification. Most related to our approach are other Bayesian decision theoretic approaches. \citet{morita2017bayesian} design a utility function $U(G, \theta)$ to identify subgroups with large treatment effects, and a review of other Bayesian developments is given by \citet{nugent2019bayesian}. The Bayesian decision theoretic approach is agnostic to the choice of model, and in particular we can use Bayesian tree-based models for inferring heterogeneous treatment effects \citep{hill2011bayesian, hahn2020bayesian}. Like this work, \citet{sivaganesan2017subgroup} use Bayesian additive regression trees (BART, \citealp{chipman2010bart}) for this task. Rather than basing decisions on the Bayes estimator of the optimal subgroup, an alternate approach is to perform uncertainty quantification on the population-level optimal $G^\star$ itself. This leads to the \emph{credible subsets} approach of \citet{schnell2016bayesian}.

\citet{foster2011subgroup} introduced the \emph{virtual twins} (VT) approach. VT begins by estimating the individual-level treatment effects using random forests and then fits a decision tree as a second-stage regression/classification algorithm to construct subgroups. Similarly, the CART algorithm has been combined with \emph{Bayesian causal forests} \citep[BCF,][]{hahn2020bayesian} to produce interpretable subgroups; for specific examples, see \citet{hahn2020bayesian} or \citet{ting2023estimating}. One of our contributions is that the BRAIDS utility exactly recovers these procedures and embeds them within a larger class of utilities with different behavior.

In the econometrics literature, policy estimation is usually framed in terms of maximizing the welfare of a population \citep{manski2004statistical}, with the \emph{empirical welfare maximization} approach selecting $G$ to maximize an empirical utility $\sum_i U_i(G, \widehat \theta)$. In randomized trials, a simple method for performing valid inference on adaptively selected subgroups is to use \emph{data splitting} as proposed by \citet{chernozhukov2018generic}. In recent work, \citet{huang2025distilling} introduce causal distillation trees, which similarly use a two-stage approach to subgroup detection using machine learning methods. In non-randomized studies, we further must take into account the possibility of selection bias and the need to estimate the propensity score. \citet{kitagawa2018who} studied minimax policy estimation, while \citet{athey2021policy} developed doubly robust policy estimators. However, data splitting can reduce statistical efficiency.

\section{Utility Functions for Bayesian Subgroup Detection}
\label{utility-functions-for-subgroup-detection}

To identify a set of subgroups or an optimal policy, Bayesian decision theoretic approaches start with a \emph{utility function} $U(G, \theta)$ where $\theta$ denotes a (possibly infinite-dimensional) parameter and $G$ denotes a collection of subgroups. The Bayes decision then maximizes the expected utility and is given by
\begin{math}
    \widehat G 
    = \arg \max_G R(G)
    \text{ where } 
    R(G) = \mathbb{E}\{U(G, \theta) \mid \mathcal D\}.
\end{math}
The population-level optimal choice is $G^\star = \arg \max_G U(G, \theta_0)$ where $\theta_0$ denotes the true value of the parameter. The utility $U(G, \theta)$ encodes what we value in a discovered subgroup. For example, \citet{morita2017bayesian} use $U(G, \theta) = \{\tau_G - \delta\} \times \frac{|N_G + 1|^\phi}{(|J| + 1)^\zeta}$, where $N_G$ is the subgroup size, $J$ is the number of defining predictors, and $\delta$, $\phi$, and $\zeta$ are tuning parameters governing the treatment-effect threshold, preference for larger subgroups, and penalty for subgroup complexity, respectively.

In this section, we will carefully construct utility functions $U(G, \theta)$ such that we have high treatment effect heterogeneity across the discovered subgroups $G = \{G_1, \ldots, G_K\}$ while keeping $G$ interpretable. In order to quantify the complexity of the partition $G$, we introduce a parameter $\vartheta$ that describes the partition structure and a mapping $G_\vartheta(x)$ such that $X_i \in G_k$ if $G_\vartheta(X_i) = k$. For concreteness, in this work we will take $G_\vartheta(x)$ to be a \emph{decision tree} with $\vartheta = \mathcal T$ where $\mathcal T$ represents the topology of the tree. The constraint that $G$ must be expressible as a decision tree greatly increases interpretability. We will consider \emph{penalized utility functions}
\begin{math}
    U(\vartheta, \theta) = U(G_\vartheta, \theta) + Q(\vartheta)
\end{math}
where $Q(\vartheta)$ penalizes the complexity of $\vartheta$ while $U(G_\vartheta, \theta)$ encodes our preference for treatment effect heterogeneity. 
For decision trees, it is natural to consider $Q(\mathcal T) = -\lambda \, \text{Depth}(\mathcal T)$ so that we consider decision trees $\mathcal T$ that can be described by a small number of splitting rules.

\paragraph{Other Parameterized Partitions}
An alternative to decision trees is to use \emph{rule lists} \citep{letham2015interpretable}, which express membership in an equivalence class $G_k$ in terms of logical rules. For example, we might partition individuals according to whether the statement $[\texttt{race} = \texttt{black} \ \text{AND} \ \texttt{age} > 70]$ is true or not. This partition is not attainable as a binary decision tree where the decision rules depend on only one predictor; \tony{consequently, rule lists are more flexible than decision trees in the sense that some partitions can be expressed with rule lists that cannot be expressed in terms of decision trees.}

\paragraph{Criticisms and Benefits of Bayesian Subgroup Detection}
Bayesian methods do not explicitly account for ``using the data twice,'' with the data used both to \emph{identify} the subgroups and to \emph{estimate} the treatment effects within subgroups.
While some have argued that this attribute of Bayesian inference is positive \citep{woody2021model}, it is natural to be uneasy about this. We study the extent to which this is an issue for our methods, and argue that regularizing the treatment effects can mostly mitigate the double-dipping behavior of Bayes estimators.
The fact that shrinkage can be used to balance model selection in inference has been seen in other contexts \citep{lockhart2014significance}. 
The payoff of the Bayesian approach is that we have higher power to detect differences because we use the full sample for inference.


\subsection{Utilities for Treatment Effect Heterogeneity}
\label{treatment-effect-heterogeneity}

A natural choice of utility that prioritizes treatment effect heterogeneity is
\begin{align}
    \label{eq:naive-loss}
    U(G, \theta) = 
    \frac{1}{N} \sum_k \sum_{i: X_i \in G_k} \{\tau(G_k) - \tau(\mathcal X)\}^2 
\end{align}
so that we are seeking the partition that maximizes how different the subgroup-level causal effects are from the average causal effect. For reasons that will become apparent, this is a \emph{risk seeking} (RS) utility. The posterior expected utility for this  $U(G, \theta)$ is given below.

\begin{theorem}[Posterior Expectation of RS Utility]
\label{prop:RS}
Under the utility function \eqref{eq:naive-loss}, the expected utility $R(G) = \mathbb{E}\{U(G, \theta) \mid \mathcal D\}$ is
given by
\begin{align}
  \label{eq:heterogeneity}
  R(G)
    &= \textnormal{const}(\mathcal D)
    + \frac{1}{N} \sum_{k = 1}^K 
    \sum_{i: X_i \in G_k}
    \operatorname{Var}\{\tau(G_k) \mid \mathcal D\}
    -\{\widehat \tau(X_i) - \widehat \tau(G_k)\}^2
\end{align}
where $\widehat \tau(G) = \mathbb{E}\{\tau(G) \mid \mathcal D\}$ and $\mathrm{const}(\mathcal D)$ is a constant independent of $G$.
\end{theorem}

The form of the expected utility in \eqref{eq:heterogeneity} is counterintuitive in that we have higher expected utility when $\operatorname{Var}\{\tau(G_k) \mid \mathcal D\}$ is \emph{large}. That is, all other things being equal, we would prefer to choose the $G_k$'s such that we are \emph{less} able to estimate the $\tau(G_k)$'s precisely. We refer to this as \emph{risk seeking} behavior. Risk seeking behavior appears to go against the likely workflow of subgroup discovery: we identify likely heterogeneous subgroups, and plan to validate these subgroups in future studies. Risk seeking behavior, by preferring higher posterior uncertainty in the $\tau(G_k)$'s, makes it less likely that subsequent studies will replicate.

We can construct a utility function that is instead \emph{risk averse} by accounting for inaccuracy in our estimates in a subsequent study. Consider the following workflow:
\begin{enumerate}
\def\labelenumi{\arabic{enumi}.}
\item
  We conduct an initial experiment to estimate the treatment effect \(\tau(\mathcal X) = \frac{1}{N} \sum_i \tau(X_i)\) and, as a secondary analysis, we produce subgroups $G_k$ as well as predictions $t_k$ for the effect within each subgroup.
\item
  Based on the recommended subgroups, a follow-up study will be performed to verify the treatment effect estimates within each group, which we assume  (for simplicity) recovers $\tau(G_k)$ without error. We then evaluate our performance on both the subgroup mean estimation and how heterogeneous the effects are across subgroups.
\end{enumerate}
A utility that captures this scenario is
\begin{align}
  \label{eq:multi-stage-loss}
  U(G, t, \theta) &= \frac{1}{N} \sum_{k=1}^K \sum_{i : X_i \in G_k} \{\tau(G_k) - \tau(\mathcal X)\}^2 
   - \lambda \frac{1}{N} \sum_{k=1}^K \sum_{i : X_i \in G_k} \{\tau(G_k) - t_k\}^2
\end{align}
where $\lambda$ is a tuning parameter used to balance the importance of finding heterogeneous subgroups on the one hand and being able to estimate the parameters on the other. We refer to \eqref{eq:multi-stage-loss} as the BRAIDS (Bayesian Risk-Aware Inference and Detection of Subgroups) utility,  a multi-stage construction motivated by considering both the current study and a hypothetical follow-up study. Below, we give the expected utility function.

\begin{theorem}[Posterior Expectation of the BRAIDS Utility]
    \label{prop:multi-stage}
    Under the utility function \eqref{eq:multi-stage-loss}, the expected utility is given by
    \begin{align}
        \label{eq:two-step-heterogeneity}
        \begin{split}
        \frac{1}{N}\sum_{k = 1}^K \sum_{i : X_i \in G_k} 
           (1 - \lambda) \operatorname{Var}\{\tau(G_k) \mid \mathcal D\}  - \lambda \{\widehat \tau(G_k) - t_k\}^2 -\{\widehat \tau(X_i) - \widehat \tau(G_k)\}^2,
        \end{split}
    \end{align}
    up to a constant. This is maximized in $(t_1, \ldots, t_K)$ when $t_k = \widehat \tau(G_k)$, yielding
    \begin{align}
        \label{eq:two-step-heterogeneity-2}
        R(G) &= \textnormal{const}(\Data) + 
        \frac{1}{N} \sum_{k=1}^K \sum_{i : X_i \in G_k}
        (1 - \lambda) \operatorname{Var}\{\tau(G_k) \mid \mathcal D\}
         -\{\widehat \tau(X_i) - \widehat \tau(G_k)\}^2.
    \end{align}
\end{theorem}

The BRAIDS utility allows us to interpolate between risk seeking behavior ($\lambda < 1$), risk neutral behavior ($\lambda = 1$), and risk averse behavior ($\lambda > 1$), with the tuning parameter $\lambda$ determining how we weight the goals of finding heterogeneity and being able to produce stable estimates. \tony{For an exploratory secondary analysis meant to generate hypotheses, a risk-neutral or slightly risk-seeking analysis may be reasonable. If the subgroups will be used to motivate a costly follow-up trial, stratify patients, or support treatment recommendations, then greater risk aversion may be preferable. In the canagliflozin case study we report results across $\lambda \in \{0,1,2\}$ as a sensitivity analysis, with $\lambda = 1$ being a pragmatic default. 
}


\paragraph{On the Choice of $K$}
\tony{Rather than thinking of $K$ as a parameter to be estimated, we view the choice of $K$ or the tree depth as defining the complexity of the interpretable approximation we are willing to report and the precision with which we can estimate the effects in each subgroup. If the tree is too shallow, the resulting estimand is a coarser best approximation to the underlying treatment-effect surface; if the tree is too deep, the subgroups may be less stable and not support precise inference on the subgroup effects.}

\subsection{Covariate Homogeneity: Why Does Risk‑Seeking Occur?}\label{covariate-homogeneity}

\tony{
  Asymptotically, the expected utility is governed by the estimated amount of heterogeneity $\sum_{k, i : X_i \in G_k} \{\widehat \tau(G_k) - \widehat \tau(\sX)\}^2$, with $\Var\{\tau(G_k) \mid \Data\}$ decreasing with the sample size. Practically, the RS behavior acts to break ties between partitions with similar degrees of estimated heterogeneity.
}
The RS behavior implied by \eqref{eq:naive-loss} is puzzling: why should an optimal decision favor subgroups whose effects we estimate \emph{less} precisely?

We argue here that the risk-seeking/risk-averse behaviors can also be interpreted as preferring \emph{covariate homogeneity} versus \emph{covariate diversity} within the discovered subgroups. Essentially, the term $\Var\{\tau(G_k) \mid \Data\}$ acts to promote covariate homogeneity within subgroups. If the $X_i$'s within a group are highly similar then the $\tau(X_i)$'s will be highly correlated in the posterior. In the extreme case where all of the $X_i$'s are exactly the same, then $\Var\{\tau(G_k) \mid \Data\} \approx \Var\{\tau(X_i) \mid \Data\}$. At the other extreme, if the $\tau(X_i)$'s are uncorrelated then $\Var\{\tau(G_k) \mid \Data\} \approx \Var\{\tau(X_i) \mid \Data\} / N_k$ where $N_k$ is the number of observations in $G_k$, which decreases with the subgroup size rather than being constant. By contrast, risk-averse utilities 
pool \emph{dissimilar} $X_i$'s in order to stabilize the average of the $\tau(X_i)$'s.

We do not believe that either of these behaviors is inherently superior. If $\operatorname{Var}\{\tau(G_k) \mid \mathcal D\}$ is sufficiently small, it may be preferable to have the subgroups be as homogeneous as possible with respect to all of the covariates. On the other hand, if statistical power is of concern, it may be more important to prioritize keeping $\operatorname{Var}\{\tau(G_k) \mid \mathcal D\}$ as small as possible. Absent domain‑specific preferences, the risk‑neutral choice \(\lambda=1\) offers a compromise.

\subsection{Algorithms and Computational Intractability}
\label{sec:algorithms}

\begin{new}

Computing optimal subgroups or treatment assignment policies requires optimizing the function $R(\vartheta) = R(G_\vartheta) + Q(\vartheta)$ over $\vartheta$, which can be very computationally challenging. We focus on the use of decision trees, and consider a penalty of the form $Q(\Tree) = -\infty$ if the depth exceeds some $d$ and $Q(\Tree) = 0$ otherwise; equivalently, we are restricting attention to only trees of depth at most $d$.

\end{new}

The most computationally favorable situation is the risk neutral setting $\lambda = 1$ of \eqref{eq:two-step-heterogeneity-2}, which is equivalent to minimizing $\sum_{i, X_i \in G_k} \{\widehat \tau(X_i) - \widehat \tau(G_k)\}^2$.  Because $\lambda = 1$ removes the variance term, we can use fast algorithms for evaluating many different splitting rules of a candidate decision tree by sharing computations across the different splitting rules \citep[see][]{fayyad1992handling}.
Despite this, optimizing $R(G)$ is NP-hard in the worst case
\citep{hyafil1976constructing}.
For depths beyond $d = 3$, we will settle for an approximate solution. We use either an evolutionary algorithm as implemented in the \texttt{evtree} package \citep{grubinger2014evtree} or the hybrid algorithm proposed by \citet{sverdrup2020policytree}.

\begin{new}

The risk averse and risk seeking settings are more challenging because we need to evaluate $\Var\{\tau(G_k) \mid \Data\}$ efficiently across many different candidate splits. To address this, we prove in the Supplementary Material that, given $S$ samples from the posterior, this term can be evaluated using the same type of cumulative-sum calculations that make the risk neutral setting fast. Specifically, if $\tau_i^{(s)}$ denotes the 
$s^{\text{th}}$ posterior draw of $\tau(X_i)$, and
\begin{math}
    \widehat \tau_i = \frac{1}{S}\sum_{s=1}^S \tau_i^{(s)},
\end{math}
then for a candidate split of a node $A$ into children $L$ and $R$, the gain in the BRAIDS objective is
\begin{math}
  \Delta_\lambda(L,R;A)
  =
  \frac{N_L N_R}{N N_A}
  \left[
    \{\widehat\tau(L)-\widehat\tau(R)\}^2
    +
    (1-\lambda)
    \left\|
      \bar r_L - \bar r_R
    \right\|_2^2
  \right],
\end{math}
where $r_i = (S - 1)^{-1/2} (\tau^{(1)}_i - \widehat \tau_i, \ldots, \tau^{(S)}_i - \widehat \tau_i)^\top$, $\bar r_B$ is the sample mean of the $r_i$'s in node $B$, and $N_B$ is the number of observations in node $B$. Because of this, after sorting the observations in a node by a candidate splitting variable, all possible thresholds can be evaluated by updating moments of $\widehat \tau_i$ and $r_i$.
This allows us to use the same strategies as for the risk-neutral setting.

\end{new}

\begin{new}
\paragraph{Bayesian Post-Selection Validity Does Not Require The Optimum}
We note that inferences we obtain conditional on a given set of subgroups are Bayes-valid regardless of how those subgroups are chosen, in the sense that nominal 95\% credible intervals will have 95\% coverage when averaged over both the prior and data. In particular, inferences reported from the posterior distribution are valid even if we use an approximation to the Bayes-optimal policy. While optimal subgroups are desirable, decision trees are notoriously unstable: slight perturbations of the data can lead to drastically different tree structures \citep{li2002instability}. In the canagliflozin trial we will see that the
  ``hybrid'' algorithm \citep{sverdrup2020policytree} performs well 
  relative to an exact algorithm, suggesting that computational approximations may be adequate for most practical purposes.
\end{new}

\begin{tocheck}

\end{tocheck}

\section{Regularization Priors and Post-Selection Inference}
\label{shrinkage-priors}

The fully Bayesian decision theoretic approach proceeds in two steps: in the first step, we fit a Bayesian model to obtain the posterior distribution of the $\tau(X_i)$'s, while in the second step we post-process the posterior to obtain subgroups and perform inferences within those subgroups. Importantly, both stages use the full dataset rather than relying on sample splitting.
A concern is that this ``double dips,'' using the data twice. The reason that double dipping can produce dishonest inference is the winner's curse --- conditional on having selected a given subgroup for inference, it is likely that we have overestimated how different it is from the average treatment effect. This produces misleading Frequentist inference. From a Bayesian perspective, however, the selection process does not matter: the data have been used only once, in the update from the prior to the posterior, and reporting posterior inferences is simply summarizing this posterior distribution \citep{woody2021model}.

We argue that if one wants to proceed in the Bayesian decision theoretic framework, it is \emph{essential} to heavily and appropriately regularize the degree of treatment effect heterogeneity. In Section~\ref{sec:post-selection-inference} we will see empirically that Bayesian inference with flat priors gives interval estimates that are badly calibrated (see Figure~\ref{fig:post-selection}) but we will also see that Bayes estimates generally perform well when an appropriate prior is used.

To reconcile the differences in performance, we note that Bayesian intervals are guaranteed to attain nominal coverage levels \emph{marginally} for $\theta$'s \emph{sampled from the prior}. Let $E$ denote, for example, the event that a posterior credible interval for the treatment effect in a data-dependent subgroup containing a particular individual covers the true subgroup effect, and suppose that our procedure is such that $\Pi(E \mid \mathcal D) = 1 - \alpha$, where $\Pi(\cdot \mid \Data)$ denotes the posterior. Then
\begin{align*}
    \Pr(E) 
    &= \int \Pr(E \mid \theta) \, \pi(\theta) \ d\theta = \int \Pi(E \mid \mathcal D) \, m(\mathcal D) \ d\mathcal D
    = (1 - \alpha) \, \int m(\mathcal D) \ d\mathcal D
    = 1 - \alpha
\end{align*}
where $m(\mathcal D)$ is the marginal distribution of the data under the prior. It follows from the above identity that it must be the case that there exist $\theta_0$'s for which $\Pr(E \mid \theta_0) \ge 1 - \alpha$. This suggests that if the true $\theta_0$ looks like a ``typical'' draw of $\theta \sim \pi(\theta)$ then we should expect the Bayesian approach to have Frequentist coverage at or above the nominal level, regardless of the fact that the subgroups are data-dependent.

Why, then, does failing to correct for post-selection inference risk the winner's curse? In our experiments, we demonstrate that when the flat-but-proper prior $\beta_\tau \sim \mathrm{Normal}(0, 100^2 \, \mathrm{I})$ is used on the regression coefficients, the winner's curse indeed occurs. The fundamental issue is a mismatch between our prior beliefs and the typical structure of treatment effects in real-world settings. Under such a flat prior, we implicitly express the wildly unrealistic belief that treatment effect heterogeneity will be massive. However, in practice, we typically expect treatment effects to be modest in magnitude, and such modest effects are \emph{not} representative draws from a flat prior. By using a regularization prior instead, we can appropriately regularize $\tau(x)$ towards homogeneity, which better reflects what is seen in both clinical trials and observational studies where dramatic differences in treatment effects across subgroups are rare. A related point concerning multiple testing was made by \citet{gelman2012we}. 

With these issues in mind, we discuss in this section how to design prior distributions that encode reasonable prior beliefs about treatment effect heterogeneity in 
practice.

\subsection{A Regularization Prior for Linear Regression}

A simple parametric model for inferring treatment effect heterogeneity is a linear model:
\begin{align}
    \label{eq:ridge}
    Y_i = \beta_{0\mu} + X_i^\top \beta_{\mu} + A_i (\beta_{0\tau} + X_i^\top \beta_\tau) + 
    \epsilon_i, \quad \epsilon_i \stackrel{\text{iid}}{\sim} \text{Normal}(0, \sigma^2).
\end{align}
Under this model, $\tau(x) = \beta_{0\tau} + x^\top \beta_{\tau}$. This parameterization allows us to separately regularize three distinct components of the model: (i) the \emph{prognostic effect} of the covariates for untreated individuals represented by $\beta_{0\mu} + x^\top \beta_\mu$; (ii) the average treatment effect, which is $\beta_{0\tau}$ when the covariates are centered; and (iii) the heterogeneity of the treatment effect, which is determined by the coefficient vector $\beta_\tau$.

For our illustrations, we use flat priors on the intercepts $(\beta_{0\mu}, \beta_{0\tau})$ and informative priors on the slope coefficients: $\beta_\mu \sim \text{Normal}(0, \sigma^2_\mu\mathbf{I})$ and $\beta_\tau \sim \text{Normal}(0, \sigma^2_\tau\mathbf{I})$. The hyperparameters $\sigma^2_\mu$ and $\sigma^2_\tau$ control the regularization applied to the prognostic and heterogeneity components. To learn an appropriate degree of regularization, we set $\sigma_\tau \sim \Exp(s_\tau)$.

\paragraph{Prior Specification and Scaling} We recommend centering and scaling the $Y_i$'s and $X_i$'s to have mean $0$ and variance $1$.
Letting $R$ denote the correlation matrix of $X_i$, treatment effect heterogeneity can be quantified via $H^2 = \Var(X_i^\top \beta_\tau \mid \beta_\tau) = \beta_\tau^\top R \, \beta_\tau$, which has average $\E(\beta_\tau^\top R \beta_\tau \mid \sigma_\tau) = \sigma^2_\tau \operatorname{tr}(R) = P \sigma^2_\tau$. Taking $s_\tau = O(P^{-1/2})$ therefore keeps the scale of heterogeneity invariant to the number of covariates.
In view of the relatively small number of predictors in the canagliflozin trial, we fixed $s_\tau = 1$ in our simulations and data analysis.
In observational studies, we also strongly recommend replacing $A_i$ in \eqref{eq:ridge} with $A_i - e(X_i)$ where $e(x)$ is (an estimate of) the propensity score. See \citet{hahn2018regularization, hahn2020bayesian, linero2024nonparametric, oganisian2025priors,ditraglia2025bayesian} for explanations of how this protects against ``regularization induced confounding.''

\subsection{Bayesian Causal Forests for Regularization}

To extend the ridge model \eqref{eq:ridge} to the nonparametric setting we can use a Bayesian causal forest \citep{hahn2020bayesian}. Remarkably, unlike for Bayesian linear regression, we will show that certain BCFs induce priors on the heterogeneity that do not depend on the design or dimensionality of the $X_i$'s. In our examples, we use a direct nonparametric extension of \eqref{eq:ridge}:
\begin{align}
    \label{eq:bcf}
    Y_i = \beta_0 + \beta_\mu(X_i) + A_i \{\tau_0 + \tau^\star(X_i)\} + \epsilon_i, 
    \qquad \epsilon_i \sim \operatorname{Normal}(0, \sigma^2).
\end{align}
This slightly modifies the parameterization of \citet{hahn2020bayesian} and allows us to place differing amounts of shrinkage on the overall treatment effect (which will be approximately equal to $\tau_0$) and the degree of treatment effect heterogeneity (captured by the function $\tau^\star(X_i)$). The treatment effect function for this model is $\tau(x) = \tau_0 + \tau^\star(x)$.

We model the unknown functions using Bayesian additive regression trees (BART). This sets
\begin{math}
    \beta_\mu(X_i) = \sum_{j=1}^{m_\mu} g(X_i; T_{\mu j}, M_{\mu j})
\end{math}
and
\begin{math}
    \tau^{\star}(X_i) = \sum_{j=1}^{m_\tau} g(X_i; T_{\tau j}, M_{\tau j})
\end{math}
where $g(x; T, M)$ denotes a regression tree with topology $T$ and terminal node parameters $M$. In our experiments, the prior on each tree is the default prior of \citet{chipman2010bart}, with $M_{j\ell} \sim \mathrm{Normal}(0, \sigma^2_\mu/m_\mu)$ for the prognostic model and $M_{j\ell} \sim \mathrm{Normal}(0, \sigma^2_\tau/m_\tau)$ for the treatment heterogeneity model.

\paragraph{The Prior on the Heterogeneity} We set $\sigma_\tau \sim \Exp(\text{scale} = s_\tau)$ for some appropriately chosen $s_\tau$. The choice of $s_\tau$ is critical in determining the degree of treatment effect heterogeneity, so we discuss this choice in more detail. Let $H^2 = \Var\{\tau^\star(X_i) \mid \tau^\star\}$ be the \emph{mean squared heterogeneity} of $\tau^\star(x)$. 
Interestingly, we can exactly compute the prior mean of $H^2$ for certain types of BART priors. We prove the following in the Supplementary Material.

\begin{theorem}
    \label{thm:bart-hetero}
    For the BART prior described in the Supplementary Material, we have $\E(H^2) = \sigma_\tau^2 (1 - e^{-\lambda / 3})$ where $\lambda$ is the average depth of a given leaf node under the prior.
\end{theorem}

Strictly speaking, Theorem~\ref{thm:bart-hetero} does not cover the BART priors used in practice; however, it works very well as an approximation (provided that the splitting rules of the ensemble are generated by uniformly sampling from the observed $X_i$'s in a given node as described by \citealp{chipman2010bart}). Applying this result as an approximation to the default prior with $\sigma_\tau = 1$ gives $\E(H^2) \approx 1 - e^{-0.4} \approx 0.33$, while for the default prior recommendation of \citet{hahn2020bayesian} we get $1 - e^{-0.086} \approx 0.082$, compared with 0.36 and 0.085 computed by Monte Carlo sampling from the prior.

\section{Simulations and Canagliflozin Application}

\subsection{Realized Utility of Subgroup Estimates}
\label{sec:algorithm-comparison}

We now compare Bayesian subgroup detection methods with other procedures in terms of their average realized utility $\E\{U(\widehat G)\}$ where the expectation is with respect to the data generating process and $\widehat G$ is an estimate of the optimal $G$. To construct plausible data generating mechanisms, we fit several models to data on $13,059$ individuals from the Medical Expenditure Panel Survey \citep[MEPS, see][]{cohen2009medical}, taking the treatment $A_i$ to be such that $a = 1$ if an individual reports that they smoke cigarettes and $a = 0$ if they do not. We emphasize that this is done only to generate plausible effect sizes $\tau(x)$ in a reproducible fashion, and so we are not concerned with ensuring that the ignorability condition holds; for the simulated datasets, the selection model is guaranteed to be ignorable. For our outcome, we take $Y_i$ to be a self-assessed measure of overall health; effect modifiers are given in the Supplementary Material.
After fitting these models, we generate synthetic datasets by sampling $X_i$'s from their empirical distribution and $A_i$'s randomly with $\Pr(A_i = 1) = 0.2$.

In all cases we generate $N = 1000$ observations. We set $Y_i \sim
\mathrm{Normal}\{\mu(X_i) + A_i \, \tau(X_i), 0.1^2 \widehat \sigma^2\}$, with the standard deviation of $0.1$ chosen to account for the fact that our datasets are only 10\% the size of MEPS. See the Supplementary Material for more detail on data generation, tuning parameters, and evaluation metrics. 

We consider six models to define plausible prognostic and treatment-effect functions: Bayesian ridge regression \eqref{eq:ridge}; unpenalized linear regression, using the same model with flat coefficient priors; BCFs \eqref{eq:bcf}; the \emph{Horserule BCF}, which estimates the trees using RuleFit \citep{nalenz2018tree}; causal random forests \citep{wager2018estimation}, implemented in \texttt{grf}; and the R-learner \citep{nie2021quasi}, implemented in \texttt{rlearner} with gradient boosted decision trees for all functions. Both causal random forests and the $R$-learner make use of the known propensity scores where possible.
\tony{Because the setting is nonparametric, the inclusion of the propensity
  score for the causal forest and the $R$-learner allows these methods to
  exactly debias the direct regression estimators while maintaining efficiency;
  when the propensity score is given rather than estimated, the bias of the final estimator is bounded by the product of the errors of the regression and propensity score \citep{dml}, which is $0$ when the propensity is known.}

 For each synthetic dataset, we refit the linear model, Bayesian ridge regression, causal random forest, $R$-learner, and horserule BCF estimators; the full BCF fit is used as a ground truth but is not refit in every simulation replication due to computational cost.

\begin{remark}
    The Horserule BCF procedure is a computationally efficient approximation to the BCF that obtains much better mixing on the leaf node parameters $\mu_{t\ell}$ of the decision trees and the crucial parameter $\sigma_\tau$.
\end{remark}

\begin{figure}
    \centering
    \includegraphics[width=.9\linewidth]{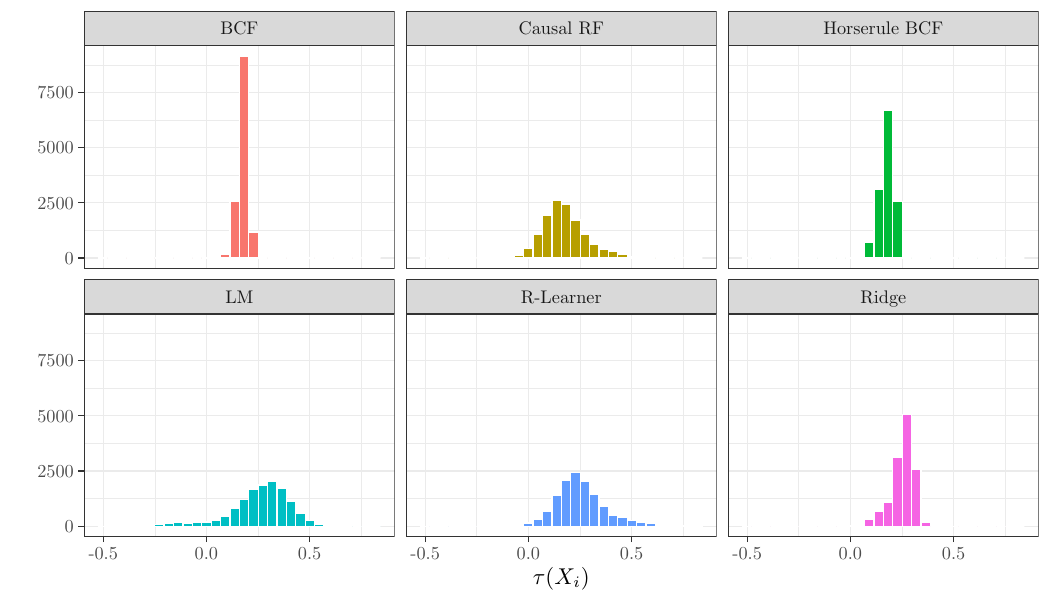}
    \caption{True values of $\tau(X)$ for the simulation in Section~\ref{sec:algorithm-comparison} for each of the data generating mechanisms we consider.}
    \label{fig:plot_regret_ptau}
\end{figure}

\begin{remark}
    Despite each of these methods being fit to the same dataset, there is a surprising amount of disagreement among the methods regarding the amount of treatment effect heterogeneity; this is displayed in Figure~\ref{fig:plot_regret_ptau}. This illustrates that different, plausible, methods for estimating heterogeneous treatment effects can easily produce very different answers regarding the degree of heterogeneity in the data. Overall, we note that all of the Bayesian approaches lead to less estimated treatment effect heterogeneity than the other methods.
\end{remark}

\paragraph{Comparison Metrics} We compare subgroups using the risk-neutral utility $U(\widehat G, \theta_0)$ as well as the mean squared error in estimating the conditional average causal effect (CATE) $\tau_0(X_i)$. For each estimator, we first obtain estimates of $\tau(X_i)$ and then the subgroup tree from these estimates.

\begin{figure}[t]
    \centering
    \hspace*{-3em}
    \includegraphics[width=1\linewidth]{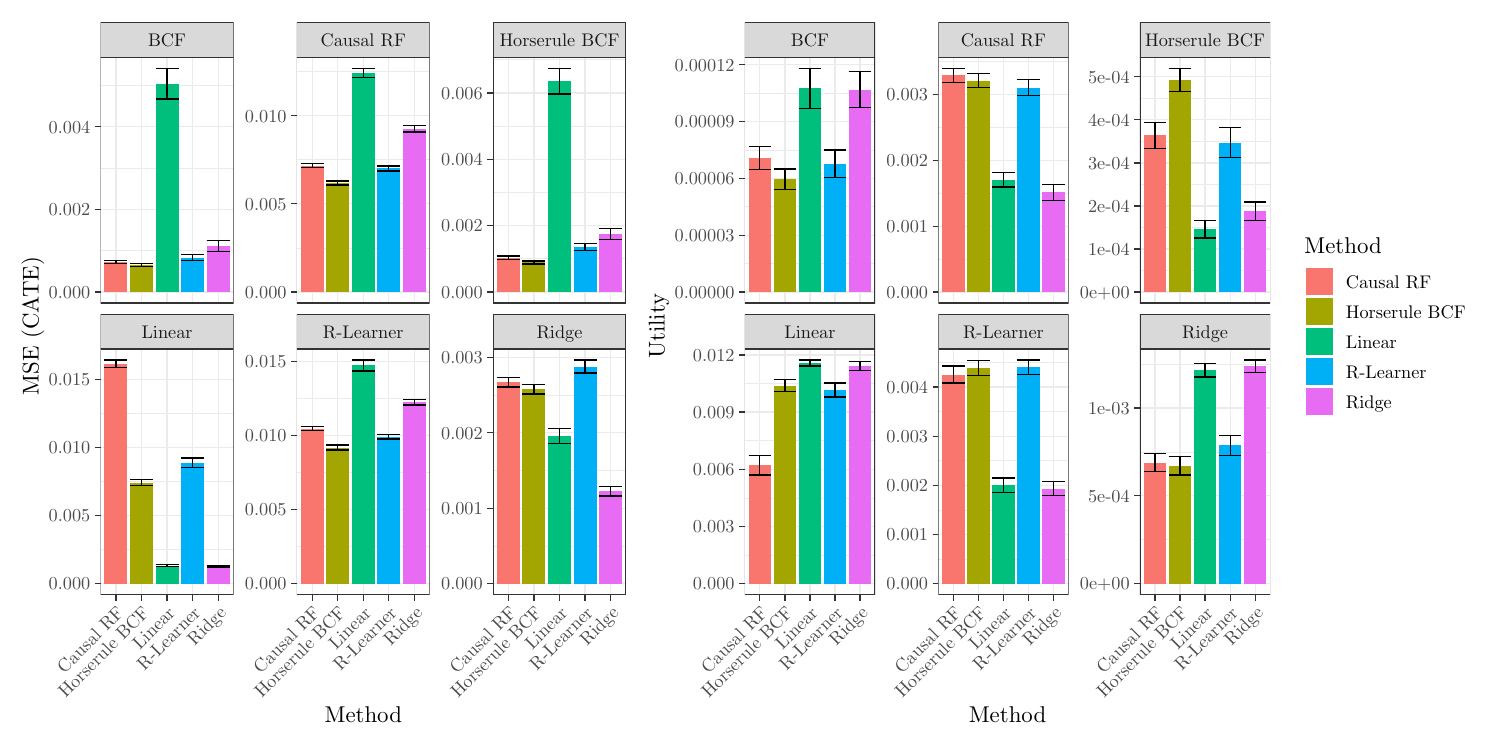}
    \caption{Results of the simulation in Section~\ref{sec:algorithm-comparison}. The MSE plots (left) display the average value of $\{\tau_0(X_i) - \widehat \tau(X_i)\}^2$ across datasets and observations. The utility plots (right) display the average utility of the discovered subgroups across the datasets. Different panels give the results for different simulation ground truths.}
    \label{fig:plot_regret}
\end{figure}

\paragraph{Conclusions} Results are given in Figure~\ref{fig:plot_regret}. When the underlying treatment effect is linear in the predictors, as expected, the non-linear methods (causal random forests, $R$-learner, and horserule BCF) perform worse than the linear methods (Bayesian ridge regression and linear regression), both in terms of having lower average utility and having higher mean squared error for the CATE. Among the non-linear methods, the horserule BCF consistently yields lower MSE than either the causal random forest or the $R$-learner across all evaluated settings, suggesting that it is comparatively more effective within the class of non-linear estimators. Additionally, the Bayesian ridge regression model we proposed dominates the unpenalized linear regression in both utility and estimation accuracy.

The situation is subtler when evaluating performance based on the identification of subgroups with the highest average utility, and the choice of estimator appears to be less critical. Some patterns still emerge: when the underlying effect is linear, linear models tend to perform better, and when the underlying model is non-linear, non-linear methods tend to perform better. Within each class of models, however, no single method consistently performs best.

\subsection{Post-Selection Inference Simulation and Double Dipping}
\label{sec:post-selection-inference}

We assess the coverage and width of nominal 95\% subgroup-effect intervals, using the BCF, ridge, and unpenalized linear models from Section~\ref{sec:algorithm-comparison} for both subgroup discovery and posterior inference. We also compare random forest and lasso outcome regressions under two strategies: ``honest'' inference on BCF-selected subgroups using an independent sample of $500$ individuals, and ``double dipping,'' where each method uses the same data to select subgroups and estimate their effects. See the Supplementary Material for additional details on the estimators and tuning parameter selection.

\paragraph{Robust estimator of subgroup effects in randomized trials} Let $N_G$ denote the number of individuals in subgroup $G$. The random forests
and lasso methods use 
\begin{math}
  \widehat \tau(G)
  = \frac{1}{N_G} \sum_{i : X_i \in G}
    \widehat \mu_1(X_i) - \widehat \mu_0(X_i) +
    \frac{\{A_i - e(X_i)\} \{Y_i - \widehat \mu_{A_i}(X_i)\}}
    {e(X_i) \{1 - e(X_i)\}}
\end{math}
to estimate the treatment effect in subgroup $G$, where $\widehat \mu_a(x)$ is
an estimate of $\mu_a(x) = \E(Y_i \mid A_i = a, X_i = x)$ and $e(x) = \Pr(A_i =
1 \mid X_i = x) = 0.2$. This estimator, which is a straightforward extension of \citet{wager2016high} to subgroup estimators, is robust when used with data splitting: due to the
fact that the propensity score $e(x)$ is known, it is immune to bias in the regression function estimator.

\paragraph{Data generation} We generate plausible data generating mechanisms in the same fashion as in Section~\ref{sec:algorithm-comparison}, fitting the BCF and ridge regression models to these data. We consider a sample size $N = 1000$ for subgroup detection, and honest methods are provided with an additional inference set of $N = 500$ individuals. In addition to taking $Y_i \sim \Normal\{\mu(X_i) + A_i \, \tau(X_i), \sigma^2\}$ where $\sigma = 0.1$, we also consider a higher-noise setting with $\sigma = 1/3$.

\begin{figure}[t]
    \centering
    \includegraphics[width=.8\linewidth]{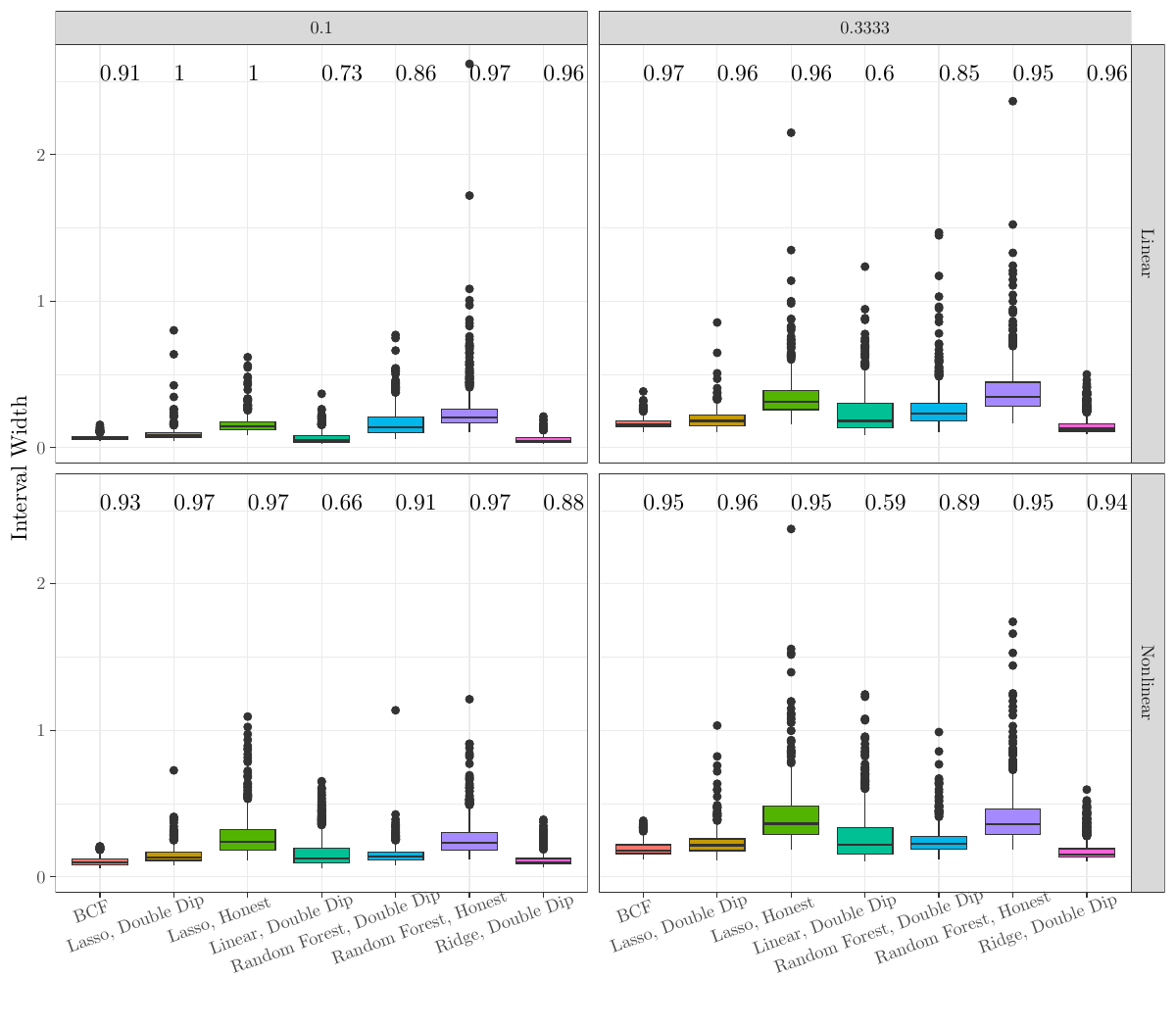}
    \caption{Results for the simulation experiment in Section~\ref{sec:post-selection-inference}. Interval width is given on the $y$-axis. Empirical coverage rates of the intervals are displayed as text next to each boxplot.}
    \label{fig:post-selection}
\end{figure}

\paragraph{Conclusions} Results for interval widths and coverage of nominal intervals are given in
Figure~\ref{fig:post-selection}, based on $200$ replications of the simulation experiment. As expected, honest methods attain or exceed the nominal coverage
level of 95\%; this is due to the use of our robust estimator of the
subgroup ATE. The downside of the honest methods is that they have access to a
smaller subset of the data for inference on the subgroup effects, and
so produce much larger intervals. Also as expected, methods that
``double dip'' without directly regularizing the treatment effects perform poorly
due to the winner's curse. This is particularly the case for linear
regression, which has coverage far below the nominal level. Random forest
methods do provide some implicit regularization as well, but do not obtain nominal coverage.

By contrast, the Bayesian approaches generally work well provided that the
regression models are well specified, and conveniently work best in the higher-noise settings representative of the original data. In this sense, the BCF performance is
somewhat more reliable, performing well in both linear and nonlinear settings.
The payoff of the Bayesian approaches is also evident: even compared to other
methods that double dip, the Bayesian methods generally produced the narrowest
intervals.

Surprisingly, the lasso appears to perform well even when double dipping, producing relatively narrow intervals and conservative inferences in all of the settings we examined. This
suggests that the simple procedure of applying the lasso to identify subgroups
and then using the same lasso fit in the robust estimator of $\widehat \tau_G$
may perform reasonably in practice.

\begin{added}
  \subsection{Post-Selection Inference and the Role of the Prior}

  We now perform a simulation study to demonstrate that (i) the frequentist
  validity of post-selection inference depends critically on the choice of a prior
  distribution that concentrates on ``plausible'' degrees of treatment effect
  heterogeneity and (ii) adopting a sensible prior on the degree of
  heterogeneity, as we generally recommend, can lead to strong coverage over a
  wide range of plausible data generating mechanisms.

\paragraph{Data Generating Process}
For each simulated dataset, we generate $N = 600$ observations with $P = 8$
covariates $X_i \sim \Normal(0, \Sigma)$ where $\Sigma_{jk} = 0.3^{|j - k|}$.
Outcomes were generated from \eqref{eq:ridge} with $\sigma^2 = 1$, $\beta_{0\mu}
= 0$, and $\beta_\mu = (0.7, -0.55, 0.4, 0.25, 0, \ldots, 0)^\top$ so that four
of the covariates are active. Treatment effect heterogeneity was then obtained
by setting $\beta_{0\tau} = 0$ and
$\beta_\tau = c (1, -0.8, 0.6, 0.35, 0, \ldots, 0)^\top$ where $c$ is chosen
so that $\Var(X_i^\top \beta_\tau) = \Delta^2$ for some $\Delta \in \{0, 0.125,
0.25, 0.375, 0.5, 0.625, 0.75, 0.875, 1, 1.5, 2\}$; this spans a range from
very low signal-to-noise ratios to very high signal-to-noise ratios.

\paragraph{Model Fitting}
For each generated dataset, we fit the corresponding Bayesian linear model with
$\beta_{0\mu}, \beta_{\mu}$, $\beta_{0\tau}$, and $\sigma$ all given flat
priors. We then set $\beta_\tau \sim \Normal(0, \sigma^2_\tau \Identity)$ with
$\sigma_\tau$ varied. If our hypothesis about the behavior of post-selection
inference is correct then we would expect the coverage of subgroup
effects to be best when $\sigma_\tau$ is matched well with the true value of
$\beta_\tau$; if $\sigma_\tau$ is chosen too large then we expect to see poor
coverage due to the winner's curse when $\Delta$ is small, while if
$\sigma_\tau$ is chosen too small when $\Delta$ is large then we expect to see
poor coverage due to an overly informative prior. In addition to varying
$\sigma_\tau \in \{0.05, 0.15, 0.35, 0.75, 1.5, 3, 20\}$ we also consider the
prior $\sigma_\tau \sim \Normal_+(0, 2^2)$ to allow the model to learn an
appropriate amount of shrinkage to apply.

\paragraph{Results}
For each simulated dataset, we computed the risk-neutral estimate of the optimal
subgroups at depths $d \in \{1, 2, 3\}$ and then, using the same fitted model,
computed whether a nominal 95\% credible interval based on the model captured
the true subgroup effect as well as the length of the credible interval. Results
based on 400 simulated datasets are given in Figure~\ref{fig:post-selection-lm}.
These results support the behavior we hypothesized, with flat priors leading to
coverage well below the nominal level for small treatment effect heterogeneity
and informative priors performing well with small treatment effect heterogeneity
but poorly when treatment effect heterogeneity is large. The best approach by
far is the hierarchical Bayesian approach with $\sigma_\tau \sim
\Normal_+(0,2^2)$, which attains nominal coverage across the full range of
$\Delta$. For small values of $\Delta$, more informative priors have narrower
intervals, but these become wider as $\Delta$ increases, to the point that they
exceed the width of intervals under the flat prior; again, the best balance comes from the
hierarchical Bayesian approach, which has an interval width that grows
appropriately with $\Delta$.

\begin{figure}[t]
  \centering
  \includegraphics[width=.8\textwidth]{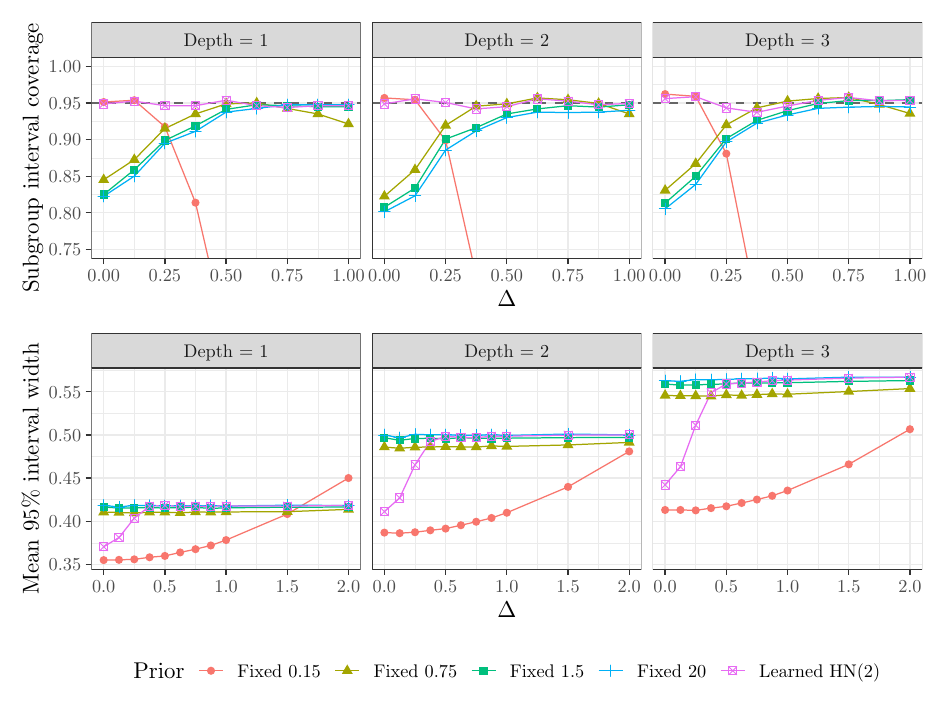}
  \caption{Top: empirical coverage of nominal 95\% subgroup treatment-effect
    intervals as the true heterogeneity parameter $\Delta$ varies, with the
    dashed horizontal line marking the nominal 0.95 target. Bottom: mean 95\%
    interval width.\label{fig:post-selection-lm}}
\end{figure}

\paragraph{Summary}
These results support our general recommendation that (i) it is important to
carefully consider the degree of treatment effect heterogeneity that is expected
if one wishes to reliably draw post-selection inferences from Bayesian
approaches and (ii) taking a hierarchical Bayesian approach that places a
weakly informative prior on $\sigma_\tau$ is a good strategy, as it can provide
good performance at many different levels of heterogeneity. The latter is
especially important, as while one generally expects treatment effect
heterogeneity to be weak, one can rarely specify it precisely a priori.
\end{added}

\subsection{Canagliflozin Clinical Trial}

We will now perform subgroup detection and inference on data from our
canagliflozin clinical trial \tony{using the Bayesian causal forest as our
  model; in the Supplementary Material, we present a comparison with Bayesian
  linear regression}. Canagliflozin is a sodium-glucose co-transporter 2 (SGLT2) inhibitor that was examined by the CANVAS program and found to reduce glycemia, blood pressure, body weight, and albuminuria in people with diabetes \citep{neal:17, perkovic:18}. 

\paragraph{Prespecified Subgroups} We first evaluate a set of prespecified subgroups of interest, as well as their interactions, using the risk-seeking, risk-neutral, and risk-averse utilities ($\lambda = 0, 1, 2$ respectively). We consider age (under or over 65), race (White, Asian, or Other), sex (Male or Female) and baseline HBA$_{1c}$ (less than 8, between 8 and 9, and higher than 9) as our subgroups. The expected utilities of the different subgroups are given in Table~\ref{tab:lambda_estimates}. Subgroup rankings are also given, with the best subgroup labeled (1) and the worst labeled (10). The subgroup rankings are relatively stable in this case across different values of $\lambda$, although we do see some notable differences. While the Race $\times$ Baseline HBA$_{1c}$ subgroup is preferred under the risk-seeking and risk-neutral settings, the Sex $\times$ Baseline HBA$_{1c}$ subgroup is preferred under a risk-averse setting; this is because race is unbalanced across groups, with 384 out of 548 individuals in our sample being White, so that estimates of differences across race groups are less precise than differences across sex.

\begin{table}[t]
\centering
\begin{tabular}{lrrr}
\toprule
\textbf{Variable}              & $\lambda=0$ & $\lambda=1$ & $\lambda=2$ \\
\midrule
Age                             &  $-2.56$ (10)  &  $-3.26$ (10)  &  $-3.96$  (9) \\
Sex                             &  $-2.25$ (8)  &  $-3.05$  (8) &  $-3.84$  (7)  \\
Race                            &  $-2.34$ (9)   &  $-3.20$ (9)  &  $-4.06$ (10)  \\
Baseline HBA$_{1c}$             &  $-1.31$ (5)   &  $-2.38$ (4)  &  $-3.45$ (4)  \\
Age $\times$ Sex               &  $-1.53$ (6)  &  $-2.55$ (6)  &  $-3.56$  (5) \\
Age $\times$ Race              &  $-1.78$ (7)   &  $-2.85$  (7)  &  $-3.91$ (8)  \\
Age $\times$ Baseline HBA$_{1c}$       &  $-0.73$ (3)   &  $-2.05$ (3)  &  $-3.37$ (3)  \\
Sex $\times$ Race              &  $-1.27$ (4)   &  $-2.47$  (5) &  $-3.66$ (6)  \\
Sex $\times$ Baseline HBA$_{1c}$       &  $-0.42$ (2)  &  $-1.85$ (2)  &  $-3.27$ (1)  \\
Race $\times$ Baseline HBA$_{1c}$      &  $-0.31$ (1)  &  $-1.81$  (1) &  $-3.32$ (2)  \\
\bottomrule
\end{tabular}
\caption{Expected utilities for the variables age, sex, race, and baseline HBA$_{1c}$ (and their interactions) across different values of $\lambda$. Rankings of the variables are given in parentheses.}
\label{tab:lambda_estimates}
\end{table}

\begin{figure}[t]
    \centering
    \includegraphics[width=.8\linewidth]{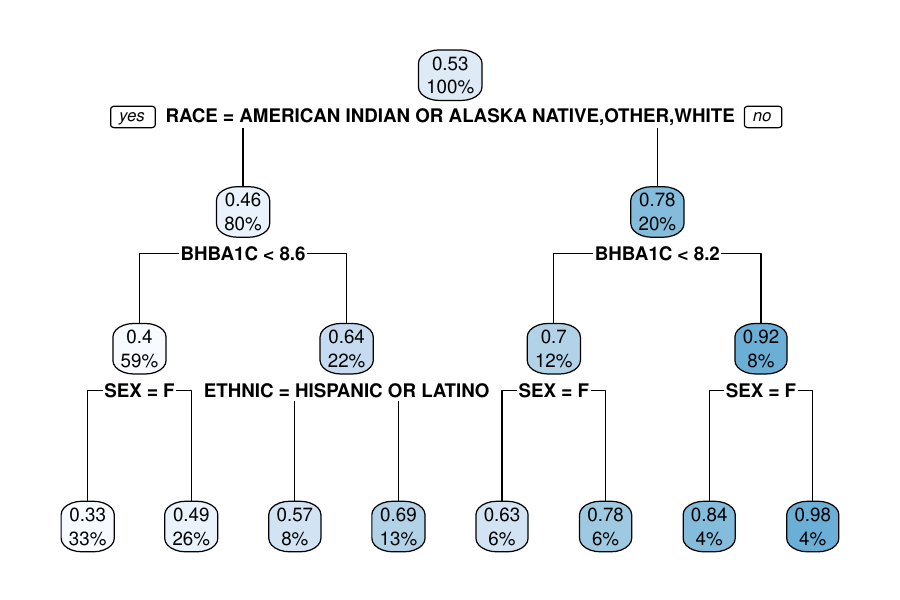}
    \caption{Posterior summarization of the treatment effect heterogeneity using the Bayesian causal forest model \eqref{eq:bcf} with $\lambda = 1$.}
    \label{fig:bart-tree}
\end{figure}

\begin{figure}[t]
    \centering
    \includegraphics[width=.9\linewidth]{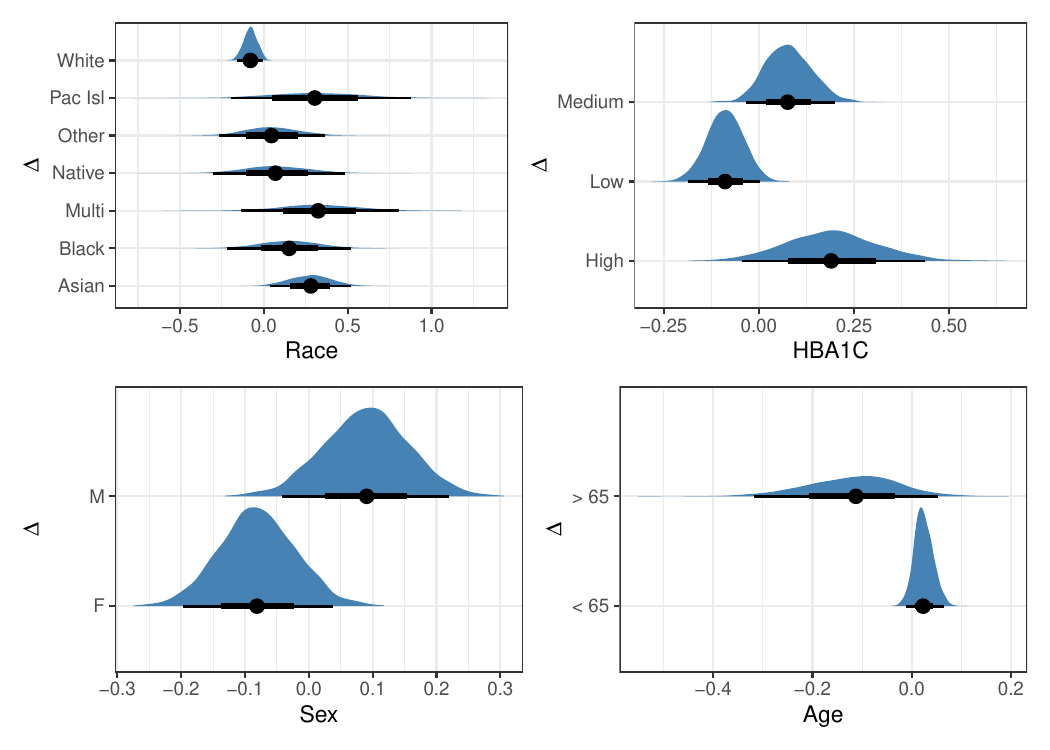}
    \caption{Posterior density for the deviation $\Delta$ from the overall average treatment effect within the prespecified subgroups defined by Race (top-left), HBA$_{1c}$ (top-right), Sex (bottom-left), or Age (bottom-right).}
    \label{fig:bart-pregroup-del}
\end{figure}

\begin{figure}[t]
    \centering
    \includegraphics[width=.8\linewidth]{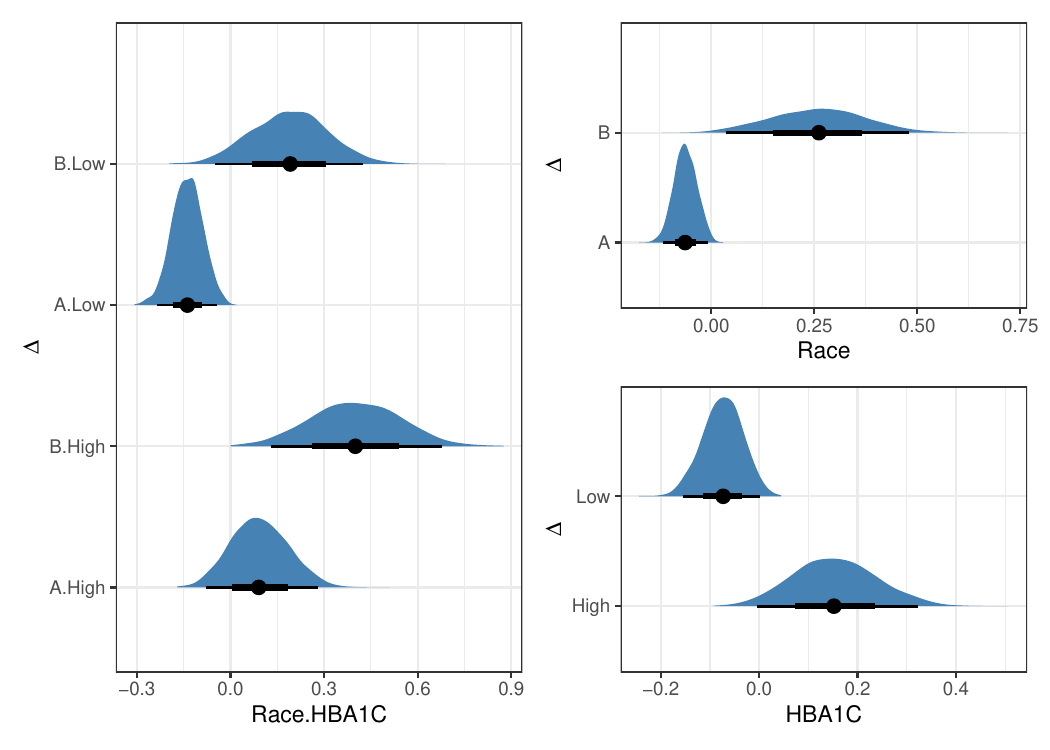}
    \caption{Posterior for the deviation $\Delta$ from the overall average treatment effect within the data-adaptive subgroups from the first- and second-level child nodes in Figure \ref{fig:bart-tree}, where the subgroup splits are based on only Race (top-right), only HBA$_{1c}$ (bottom-right), or both (left). Race is divided into group A (White, Other and Native American) and group B (Black, Asian, Pacific Islander, and Multi-Racial), while HBA$_{1c}$ is divided as $< 8.4$ or $> 8.4$.
    }
    \label{fig:bart-subgroup-del}
\end{figure}

\paragraph{Comparing Learned Subgroups to Prespecified Subgroups}
Figure \ref{fig:bart-tree} displays the subgroups discovered by the risk-neutral approach, which identifies race, baseline HBA$_{1c}$, and biological sex as the most significant treatment effect modifiers. Figure~\ref{fig:bart-pregroup-del} and Figure~\ref{fig:bart-subgroup-del} respectively show the posterior distribution of the deviations $\Delta_G = \tau_G - \tau_{\mathcal X}$ of the subgroup ATEs from the overall ATE for prespecified groups and the estimated subgroups. The results provide substantial evidence of differences between the different race subgroups and the baseline HBA$_{1c}$ subgroups, both individually and jointly. For instance, as suggested by Figure \ref{fig:bart-subgroup-del}, the estimated difference in treatment effect between race group A with low baseline HBA$_{1c}$ and race group B with high baseline HBA$_{1c}$ is  0.55, with a 95\% credible interval of (0.18, 0.91) and a posterior probability of a negative difference equal to $P = 4 \times 10^{-4}$. We note that differences across racial groups are less distinct for the prespecified groups, as aggregation across races is needed to find significant effects.

\begin{added}
  \paragraph{Adaptive Subgroup Comparisons for Different $\lambda$}

  Table~\ref{tab:ablation} shows how the results of the subgroup analysis change
  as we vary $\lambda$, the method used to find the subgroups, and the maximal
  number of subgroups $K$. The resulting decision trees are given in the
  Supplementary Material. Let $B = \sum_k \sum_{i : X_i \in G_k} \{\widehat
  \tau(G_k) - \widehat \tau(\sX)\}^2 / N$ and $V = \sum_k \sum_{i : X_i \in G_k}
  \Var\{\tau(G_k) \mid \Data\} / N$, respectively, so that $R(G) =
  \text{const}(\Data) + B + (1 - \lambda) V$; larger values of $|1 - \lambda|$ correspond to placing a larger weight on $V$. We compare different subgroup selection criteria by their values of $B$, $V$, a two-sided posterior $P$-value in the subgroup with the smallest estimated ATE (Low $P$) and the largest estimated ATE (High $P$), and the average (unweighted) standard error of the ATE differences $\Delta$ across subgroups. Methods in Table~\ref{tab:ablation} correspond to exact optimization of the utility (E), greedy CART optimization (C), a ``hybrid'' approach described by \citet{sverdrup2020policytree} (H), and prespecified subgroups from the clinical trial design (P).

\begin{table}[t]
  \centering
  \resizebox{\textwidth}{!}{
    \begin{tabular}{llllrrrrr}
      \toprule 
      $\lambda$ & Method & $K$ & Splits & $B$ & $V$ & Low $P$ & High $P$ & SE Mean\\
      \midrule
      0 & E & 4 & Race, \{Ethnic, HBA$_{1c}$\} & 289 & 130 & $8\times10^{-4}$ & $6\times10^{-3}$ & 0.1267\\
      1 & E & 4 & Race, HBA$_{1c}$ & 301 & 110 & 0.0012 & 0.0008 & 0.1224\\
      2 & E & 4 & Race, HBA$_{1c}$ & 301 & 110 & 0.0012 & 0.0008 & 0.1224\\
      0 & C & 8 & Race, HBA$_{1c}$, \{Ethnic, Sex\} & 359 & 158 & 0.0032 & 0.0004 & 0.1433\\
      1 & C & 8 & Race, HBA$_{1c}$, \{Ethnic, Sex\} & 359 & 159 & 0.0032 & 0.0004 & 0.1433\\
      2 & C & 8 & Race, HBA$_{1c}$, \{Age, HBA$_{1c}$, Sex\} & 346 & 141 & 0.0032 & 0.0012 & 0.1361\\
      0 & H & 8 & Race, \{HBA$_{1c}$, Race\}, \{Ethnic, HBA$_{1c}$\} & 383 & 174 & $4\times10^{-3}$ & $8\times10^{-4}$ & 0.1459\\
      1 & E & 8 & Race, \{HBA$_{1c}$, Race\}, \{Ethnic, HBA$_{1c}$\} & 383 & 174 & $4\times10^{-3}$ & $8\times10^{-4}$ & 0.1459\\
      2 & H & 8 & Race, \{HBA$_{1c}$, Race\}, \{Ethnic, HBA$_{1c}$, Sex\} & 360 & 159 & 0.0032 & .0004 & 0.1417\\
      --- & P & 2 & Age & 27 & 63 & 0.196 & 0.196 & 0.0922\\
      --- & P & 3 & HBA$_{1c}$ & 121 & 89 & 0.0464 & 0.0864 & 0.1030\\
      --- & P & 2 & Sex & 71 & 84 & 0.1888 & 0.1888 & 0.0917\\
      --- & P & 6 & Race & 195 & 121 & 0.0264 & 0.1768 & 0.1874\\
      --- & P & 6 & HBA$_{1c}$, Sex & 179 & 132 & 0.0272 & 0.0768 & 0.1238\\
      \bottomrule
    \end{tabular}
  }
  \caption{Comparison of different subgroup-finding procedures. The measures $B$ and $V$ are multiplied by $10^4$ for ease of reading. \label{tab:ablation}}
\end{table}

Some findings are stable across all settings. Race is always selected as the root split, and baseline HBA$_{1c}$ appears in every tree. Consequently, our primary substantive conclusions do not depend on the optimization or choice of $\lambda$. By contrast, ethnicity, sex, and age appear only in deeper splits and vary across $\lambda$. For the hybrid approaches at $K = 8$, ethnicity is also used in all trees; however, the risk-averse setting prefers to use sex for one of its splits; this is because differences across sex are more stable than differences across ethnicity.

  The changes across $\lambda$ illustrate the risk-aware behavior that motivates
  BRAIDS. At $K = 4$, the risk-seeking utility also includes ethnicity, as
  uncertainty in the size of the ethnicity effect leads to a larger $V$; both
  the risk-neutral and risk-averse utilities, by contrast, select the same
  subgroups. We also see that, regardless of method or $K$, the average
  within-group standard error decreases with $\lambda$. Interestingly, the
  risk-seeking utilities produce the lowest $P$-values in the high/low 
  subgroups.

  The data-adaptive trees also find much more pronounced heterogeneity than the
  prespecified subgroups. Even the depth-two adaptive trees have values of $B$
  at least 50\% larger than those of the prespecified subgroups. Additionally, both
  extreme leaves of every adaptive tree have posterior equal-tail $P$-values no
  larger than $0.006$. By comparison, the prespecified subgroups find much
  weaker evidence for treatment effect heterogeneity; we find some support for
  effects of sex, HBA$_{1c}$, and race, but it is more tenuous.

  Part of the value added by the adaptive subgroup estimation strategy is that,
  in this case, it pools information across races. This is important for
  identifying race effects, as many of the racial subgroups are quite small.
  Most of the heterogeneity associated with race, in fact, can be captured just by
  knowing if an individual is white or not. The prespecified subgroup analysis
  in the clinical trial did not pool races when examining subgroup effects,
  missing out on this source of possible heterogeneity.

  Finally, our main scientific conclusions are relatively insensitive to the choice of optimization method, with exact or hybrid approaches identifying subgroups with smaller $P$-values. While in this case it does not matter --- all of the data-adaptive analyses find strong evidence of heterogeneity --- it is easy to imagine more borderline cases where it would matter. On this dataset the benefit in terms of raw utility increase occurred for $\lambda = 0$, and made the least difference for $\lambda = 2$, but it is hard to say whether this speaks to a more general trend.
\end{added}

\section{Discussion}

In this paper we studied the Bayesian decision-theoretic framework for subgroup identification, with an emphasis on discovering subgroups based on treatment effectiveness and treatment effect heterogeneity. We have argued that it is essential to appropriately regularize the treatment effect function in order to safely apply Bayesian machine learning methods. We also introduced the BRAIDS utility, which establishes a continuum between risk-seeking, risk-neutral, and risk-averse subgroup selection; this makes explicit the tradeoff between maximizing treatment effect heterogeneity and maintaining stability in subgroup-level estimates. 
We show that the risk-neutral utility precisely recovers a variant of the virtual twins algorithm in \citet{foster2011subgroup}. This perspective places otherwise heuristic methods within a broader class of fully Bayesian decision rules.

A central contribution of this work is the demonstration that fully Bayesian inference can maintain near-nominal Frequentist coverage, even when subgroups are identified adaptively. Contrary to the prevailing concern that Bayesian post-selection inference must explicitly adjust for data reuse and ``double dipping'', we show empirically that appropriate prior regularization can mitigate selection-induced bias. Our results on both synthetic and real data show that, under such priors, posterior credible intervals achieve near-nominal coverage for subgroup treatment effects, while also avoiding the inefficiencies of sample splitting. This finding suggests that careful prior specification can serve as a practical alternative to sample splitting, yielding efficient inference without compromising validity.

Our approach has a few limitations. First, the validity of Bayesian post-selection inference depends on the use of well-calibrated priors and our ability to express realistic beliefs about treatment effect heterogeneity. Shrinkage priors can improve coverage, but performance deteriorates when overly diffuse priors are used. Second, while the use of decision trees enhances interpretability, their inherent instability remains a concern, especially when small perturbations in the data yield markedly different tree structures. While our results suggest that approximate optimization is often sufficient for practical purposes, exact Bayes-optimal subgroup selection is computationally challenging. We plan to address these limitations in future work. 

\paragraph{Acknowledgements}
This study, carried out under YODA Project 2024-0600, used data obtained from the Yale University Open Data Access Project, which has an agreement with JANSSEN RESEARCH \& DEVELOPMENT, L.L.C.. The interpretation and reporting of research using this data are solely the responsibility of the authors and does not necessarily represent the official views of the Yale University Open Data Access Project or JANSSEN RESEARCH \& DEVELOPMENT, L.L.C.. The original proposal can be found: https://yoda.yale.edu/data-request/2024-0600/

\paragraph{Data Availability Statement}
The data that support the findings of this study are available from the YODA Project. Restrictions apply to the availability of these data, which were used under license for this study. Data are available from \url{yoda.yale.ed} with the permission of the YODA project and Janssen Research and Developement, L.L.C.

\bibliographystyle{apalike}
\bibliography{references}

\end{document}


\title{Supplementary Material to Decision Theoretic Subgroup Detection With Bayesian Machine Learning}
\date{}

\author{
  Entejar Alam\thanks{\texttt{entejar@utexas.edu}, Equal contribution},
  Poorbita Kundu\thanks{\texttt{poorbitakundu@gmail.com}, Equal contribution},
  and Antonio R. Linero\thanks{\texttt{antonio.linero@austin.utexas.edu}}
}

\maketitle
\tableofcontents

\doublespacing

\section{Proof of Theorem~\ref{thm:bart-hetero}}

The BART prior we consider is a modification of the BART prior of \citet{chipman2010bart} in two ways:
\begin{enumerate}
    \item Rather than a branching process with branching probabilities given by $p(d) = \alpha / (1 + d)^\beta$ for a node of depth $d$, we instead choose $p(d) = \Pr(Z > d \mid Z \ge d)$ where $Z$ has a Poisson distribution with mean $\lambda$; a consequence of this is that the depth of the terminal node associated with any given $X_i$ also has a Poisson distribution with mean $\lambda$.
    \item We assume the $X_{ij}$'s are continuous random variables, and that cutpoints are sampled by (i) randomly choosing some axis $j$ and (ii) sampling the cutpoint from the distribution of $[X_{ij} \mid X_i \text { is associated with the current node}]$. 
\end{enumerate}
Given these assumptions, we start by writing $\tau^\star(x)$ as
\begin{align*}
    \tau^\star(x) = \sum_{t, \ell} A_{t\ell}(x) \ \mu_{t\ell}
\end{align*}
where $A_{t\ell}$ is the event that $x$ is associated with leaf node $\ell$ of tree $t$. Taking the variance of $X_i \sim F_X$ gives
\begin{align*}
    &\sum_{(t,\ell)} \Var\{A_{t\ell}(X_i)\} \, \mu_{t\ell}^2 + 
      2 \sum_{(t, t', \ell, \ell'): (t, \ell) \ne (t', \ell')} \Cov\{A_{t\ell}(X_i) \, A_{t'\ell'}(X_i)\} \, \mu_{t\ell} \, \mu_{t'\ell'}
    \\&\quad = \sum_{(t, \ell)} p_{t\ell} (1 - p_{t\ell}) \mu_{t\ell}^2
    - 2 \sum_{(t, t', \ell, \ell'): (t, \ell) \ne (t', \ell')} p_{t\ell} \, p_{t'\ell'} \, \mu_{t\ell} \, \mu_{t'\ell'},
\end{align*}
where $p_{t\ell}$ is the probability that $X_i$ is associated with $(t,\ell)$ when $X_i \sim F_X$. We first average out the $\mu_{t\ell}$'s, which, because they are mean $0$ and have variance $\sigma^2_\tau / m_\tau$, gives
\begin{align*}
    \frac{\sigma^2_\tau}{m_\tau} \sum_{t = 1}^{m_\tau} \sum_\ell (p_{t\ell} \, 
    - p_{t\ell}^2)
    = 
    \frac{\sigma^2_\tau}{m_\tau} \sum_{t = 1}^{m_\tau} (1 - q_{t})
\end{align*}
where $q_{t} = \sum_\ell p_{t\ell}^2$ can be interpreted as the probability that, if $X_i \sim F_X$ and $X_{i'} \sim F_X$, we observe the event that $X_i$ and $X_{i'}$ share the same leaf node in tree $t$. Because the trees are sampled iid from the same prior, averaging over the tree we get
\begin{align*}
    \E(H^2) = \sigma^2_\tau (1 - \bar q)
\end{align*}
where $\bar q$ is the probability that $X_i \sim F_X$ and $X_{i'} \sim F_X$ share the same leaf node in a randomly-sampled tree $\Tree$ from the prior distribution.

Now, consider $X_i, X_{i'} \sim F_X$ and consider growing a decision tree $\Tree$. Given that $X_i$ and $X_{i'}$ are associated with a given node $b$ then, if that node becomes a branch, the probability that they will both go right is $\Pr(X_{ij} > c \cap X_{i'j} > c)$ where $c$ is drawn from the $j$-marginal of $F_X$ restricted to node $b$; but $X_{ij}$ and $X_{i'j}$, because they are also samples from $F_X$ that are associated with node $b$, also have this distribution, so this probability is just the probability that $c$ is the smallest of three samples taken from the same continuous distribution, which is $1/3$. Similarly, the probability that both go left is also $1/3$, so the probability that $X_i$ and $X_{i'}$ remain together is $2/3$, irrespective of which $j$ is sampled to construct the split.

After $k$ splits, the probability that $X_i$ and $X_{i'}$ remain together then becomes $(2/3)^k$. But the depth of the node associated with $X_i$, $K_i$, has a $\Poisson(\lambda)$ distribution, so $\bar q = \E\{(2/3)^{K_i}\} = e^{-\lambda / 3}$. Putting all of this together, we have
\begin{align*}
    \E(H^2) = \sigma^2_\tau (1 - e^{-\lambda / 3}).
\end{align*}
We can also note a couple of other variants of this result that can be derived using the same argument:
\begin{enumerate}
    \item If instead of splitting randomly we split at the median value of the $X_{ij}$'s at a given node, we would instead get $\sigma^2_\tau (1 - e^{-\lambda / 2})$. This is because the probability that $X_i$ and $X_{i'}$ remain together is $1/2$ rather than $2/3$.
    \item If instead of using the Poisson distribution we used the \citet{chipman2010bart} prior with $\beta = 0$ and $\alpha \le 0.5$, we instead get $\sigma_\tau^2 \times \frac{\alpha}{3 - 2\alpha}$. This comes from replacing the Poisson distribution with a geometric distribution with success probability $1 - \alpha$.
\end{enumerate}

\section{Proof of Theorem~\ref{prop:multi-stage}}

We prove Theorem~\ref{prop:multi-stage}, noting that Theorem~\ref{prop:RS} is obtained as the special case with $\lambda = 0$. Taking the expected value of \eqref{eq:multi-stage-loss} with respect to the posterior gives
\begin{align*}
      R(G, t) 
      &= \frac{1}{N} \sum_{k=1}^K \sum_{i : X_i \in G_k} 
        \big[\Var\{\tau(G_k) - \tau(\mathcal X) \mid \Data\} + 
        \{\widehat \tau(G_k) - \widehat \tau(\mathcal X)\}^2
        \\&\quad - \lambda \Var\{\tau(G_k) \mid \Data\} - \lambda \{\widehat \tau(G_k) - t_k\}^2\big].
\end{align*}
We now make two observations. First, by standard ANOVA arguments, we know 
\begin{align*}
    \sum_i \{\widehat \tau(X_i) - \widehat \tau(\mathcal X)\}^2 = 
    \sum_{k, i: X_i \in G_k} \{\widehat \tau(X_i) - \widehat \tau(G_k)\}^2 + 
    \sum_{k, i: X_i \in G_k} \{\widehat \tau(G_k) - \widehat \tau(\mathcal X)\}^2.
\end{align*}
Hence, $\sum_{k, i : X_i \in G_k} \{\widehat \tau(G_k) - \widehat \tau(\mathcal X)\}^2$ can be replaced by $\text{const}(\Data) - \sum_{k, i: X_i \in G_k} \{\widehat \tau(X_i) - \widehat \tau(G_k)\}^2$. 

Our second observation is that
\begin{align*}
    &\sum_{k, i: X_i \in G_k} \Cov\{\tau(G_k), \tau(\mathcal X)\}
    \\&= \Cov\left\{\sum_{k, i} \tau(G_k), \tau(\mathcal X)\right\}
    \\&= N \Cov\left\{\tau(\mathcal X), \tau(\mathcal X)\right\}
    \\&= \sum_{k, i: X_i \in G_k} \Var\{\tau(\mathcal X)\}.
\end{align*}
Because of this, we can write
\begin{align*}
    &\sum_{k} \sum_{i : X_i \in G_k} \Var\{\tau(G_k) - \tau(\mathcal X) \mid \Data\}
    \\&= \sum_{k} \sum_{i : X_i \in G_k} \Var\{\tau(G_k) \mid \Data\} - 2 \Cov\{\tau(G_k), \tau(\mathcal X) \mid \Data\} + \Var\{\tau(\mathcal X) \mid \Data\}
    \\&= \sum_{k} \sum_{i : X_i \in G_k} \Var\{\tau(G_k) \mid \Data\} - \Var\{\tau(\mathcal X) \mid \Data\}
    \\&= \text{const}(\Data) + \sum_{k, i : X_i \in G_k} \Var\{\tau(G_k) \mid \Data\}.
\end{align*}
Putting our two observations together, we get
\begin{align*}
    R(G, t) &= \text{const}(\Data)
    \\& + \frac{1}{N}\sum_{k, i: X_i \in G_k} (1 - \lambda) \Var\{\tau(G_k) \mid \Data\}  - \{\widehat\tau(X_i) - \widehat \tau(G_k)\}^2 - \lambda \{\widehat \tau(G_k) - t_k\}^2.
\end{align*}
This expression is maximized in $t$ when all the $\{\widehat \tau(G_k) - t_k\}^2$'s are zero, i.e., $t_k = \widehat \tau(G_k)$. This gives us the final criterion
\begin{align*}
    R(G)= \text{const}(\Data) + \frac{1}{N}\sum_{k, i: X_i \in G_k} (1 - \lambda) \Var\{\tau(G_k) \mid \Data\}  - \{\widehat\tau(X_i) - \widehat \tau(G_k)\}^2
\end{align*}
as desired.

\section{Detailed Simulation Designs}
\label{sec:detailed-simulation-designs}

This section gives an explicit description of the two MEPS-based simulation experiments reported in the main text. The first experiment compares methods by realized utility and CATE estimation error. The second experiment studies post-selection inference and the effect of using the same data for subgroup discovery and subgroup effect estimation.

\subsection{Realized Utility of Subgroup Estimates}
\label{sec:supp-realized-utility-design}

The purpose of this simulation was to compare several treatment-effect
estimators in terms of the quality of the subgroups they induce. Rather than
specifying a stylized parametric ground truth, we constructed a collection of
plausible data-generating mechanisms by fitting several heterogeneous
treatment-effect methods to the Medical Expenditure Panel Survey (MEPS) data and
then treating the fitted prognostic and treatment-effect functions as the ground truth.

\paragraph{MEPS preprocessing and construction of data generating mechanisms}
We retained individuals with positive income and used self-assessed health as the outcome. The treatment indicator was \texttt{smoke}, equal to one for individuals who reported smoking cigarettes and zero otherwise. For effect modifiers we considered sex, age, log income, race, census region, health insurance status, education level, marital status, and family size.

We fit six models to these data to obtain six possible ground-truth pairs $\{\mu_0(x), \tau_0(x)\}$: an unpenalized linear model with treatment-covariate interactions, the Bayesian ridge model, a causal random forest, a Bayesian causal forest, the $R$-learner using gradient boosting with the \texttt{rboost} function, and the horserule BCF approximation.

For a fitted model $m$, we define $\tau_m(x)$ as its fitted CATE and $\mu_m(x,a)$ as its fitted conditional mean under treatment level $a$. The simulation treats these fitted functions as fixed truth. This construction is used only to obtain realistic magnitudes and shapes for $\mu(x)$ and $\tau(x)$; in the simulated datasets, treatment is randomized by construction.

\paragraph{Synthetic data generation}
For each fitted model $m$, each replication proceeds as follows.
\begin{enumerate}
\item  We sample $N = 1000$ rows from the empirical MEPS covariate distribution with replacement. 
\item We sample random treatment assignments $A_i \sim \Bernoulli(0.2)$ and outcomes $Y_i = \mu_m(X_i, A_i) + \epsilon_i$ with $\epsilon_i \sim \Normal(0, 0.1^2)$ (the standard deviation of $0.1$ for the errors is consistent with the original MEPS data).
\end{enumerate}
In this way, we constructed $200$ synthetic datasets.

\paragraph{Methods refit on each synthetic dataset}
On each synthetic dataset, we refit five estimators: the unpenalized linear model, Bayesian ridge regression, causal random forests, the $R$-learner, and horserule BCF. For causal random forests and the $R$-learner, the known randomization probability $e(x) = 0.2$ was supplied when the software interface allowed it.

\paragraph{Priors and Posterior Sampling}

For the Bayesian linear ridge regression model we set
\begin{align*}
  \beta_j \mid \sigma_\mu \sim \Normal(0,\sigma_\mu^2),\qquad
  \tau_j \mid \sigma_\tau \sim \Normal(0,\sigma_\tau^2),\qquad
  \sigma_\mu, \sigma_\tau \sim \Exp(\mathrm{rate}=r),
\end{align*}
with $r = 1$. The same prior is used for the horserule BCF after extracting tree-based features, except that we take $r = Q^{-1/2}$ where $Q$ is the number of extracted rule features. In total, $Q = 1000$ rules were extracted for the horserule procedure, using the default settings of the \texttt{horserule} package. Both the ridge and horserule BCF were fit using the \texttt{rstan} package using 4 chains with 1000 warmup iterations and 1000 collected samples. Both intercept parameters use flat $\Normal(0, 100^2)$ priors. 

\paragraph{$R$-learner tuning}
We use the \texttt{rboost} function in the \texttt{rlearner} package. With $N = 1000$ observations, it uses the default of ten-fold cross-validation with ten randomly-sampled hyperparameter configurations, at most 1000 boosting rounds, and early-stopping after 10 rounds of no improvement.
The search covers learning rate, tree depth, row and column subsampling, and split/leaf controls. This tuning is applied to the outcome-nuisance and treatment-effect fits.

\paragraph{Causal-forest settings}
We use the \texttt{causal\_forest} function in the \texttt{grf} package to estimate the CATEs, with the package defaulting the parameter \texttt{mtry} (i.e., the number of candidate variables to split on at a given node) to $P$ for $P = 18$ predictors (i.e., using the full set of variables at every split). Other defaults include 2000 trees, resampling 50\% of rows without replacement for each tree, and an ``honesty fraction'' of 50\%. 

\paragraph{Subgroup construction and metrics}

For each fitted estimator, we computed the estimated CATEs $\widehat \tau(X_i)$ and constructed a $K = 4$ subgroup regression tree under $\lambda = 1$; this is equivalent to constructing an optimal regression tree under squared-error loss with the $\widehat \tau(X_i)$ as the response and the MEPS covariates as predictors. We used the \texttt{evtree} package to construct the decision trees.

Two metrics were computed. First, the CATE mean squared error was
\begin{math}
  \frac{1}{N} \sum_{i=1}^N \{\widehat \tau(X_i)-\tau_0(X_i)\}^2.
\end{math}
Second, the realized subgroup utility was evaluated under the true treatment-effect surface as
\begin{math}
  \frac{1}{N} \sum_{k} \sum_{i:X_i \in G_k}
  \{\tau_0(G_k)-\tau_0(\mathcal X)\}^2,
\end{math}
where $\tau_0(G_k)$ is the average true CATE among observations assigned to subgroup $k$.

Algorithm~\ref{alg:realized-utility-workflow} describes the full workflow for the simulation.

\begin{algorithm}
\caption{MEPS-based realized-utility simulation workflow}
\label{alg:realized-utility-workflow}
\begin{algorithmic}[1]
\Require MEPS data, candidate truth models $\mathcal M$, estimators $\mathcal A$, replication count $B=200$
\State Preprocess MEPS covariates and construct the design matrix.
\ForAll{truth models $m \in \mathcal M$}
  \State Fit $m$ to MEPS and store fitted $\mu_m(x,a)$ and $\tau_m(x)$.
\EndFor
\ForAll{truth models $m \in \mathcal M$}
  \For{$b = 1,\ldots,B$}
    \State Sample $N=1000$ covariate rows from MEPS with replacement.
    \State Draw $A_i \sim \Bernoulli(0.2)$ and $Y_i=\mu_m(X_i,A_i)+\epsilon_i$, $\epsilon_i \sim \Normal(0,0.1^2)$.
    \ForAll{estimators $a \in \mathcal A$}
      \State Fit $a$ to the simulated dataset and compute $\widehat\tau_a(X_i)$.
      \State Fit a depth-2 regression tree to $\widehat\tau_a(X_i)$ to obtain subgroups $G_a$.
      \State Record CATE MSE and realized risk-neutral utility under the true $\tau_m(X_i)$.
    \EndFor
  \EndFor
\EndFor
\end{algorithmic}
\end{algorithm}

\subsection{Post-Selection Inference and Double Dipping}
\label{sec:supp-post-selection-design}

The second MEPS-based experiment evaluates whether nominal 95\% subgroup treatment-effect intervals cover the true subgroup effects after the subgroups have been selected adaptively. The experiment compares fully Bayesian approaches that use the same data for subgroup selection and posterior inference, frequentist procedures that use a held-out inference sample, and frequentist procedures that use the same data for both stages. The same strategy for preprocessing the data and constructing data generating mechanisms was used for this experiment as in the experiment described in Section~\ref{sec:supp-realized-utility-design}. We restrict attention to the Bayesian ridge regression data generating mechanism for the ``linear'' setting and the horserule BCF for the ``nonlinear'' setting. The lasso was fit using the \texttt{glmnet} package using the \texttt{cv.glmnet} function, which performs cross-validation over a grid.

\paragraph{Synthetic data generation}
For each replication and each ground truth, we generated an adaptive-selection training sample of size $N_{\mathrm{train}}=1000$. For methods using honest inference, we also generated an independent inference sample of size $N_{\mathrm{test}}=500$. In both samples, covariate rows were sampled from the empirical MEPS covariate distribution, treatment was randomized as
\begin{math}
  A_i \sim \Bernoulli(0.2),
\end{math}
and outcomes were generated from the fitted linear or nonlinear truth.

Both noise settings labeled $\sigma=0.1$ and $\sigma=1/3$ in the main text
are included, with 200 replications for each truth and noise setting.
More precisely, the supplied scripts use these values as noise multipliers:
with $c\in\{0.1,0.3333\}$ (the latter approximating $1/3$), they generate
\[
  Y_i=\mu_m(X_i)+A_i\tau_m(X_i)+\epsilon_i,
  \qquad \epsilon_i\sim\Normal\{0,(c\widehat\sigma_m)^2\},
\]
where $\widehat\sigma_m$ is the posterior mean residual standard deviation from fitting truth model $m$ to MEPS; $\widehat \sigma_m$ happens to be approximately 1 (specifically  $\widehat \sigma_m \approx 0.92$).

\paragraph{Methods and interval construction}

This experiment uses similar priors for the ridge and horserule BCF as before, except that the prognostic coefficients are given weakly-informative $\Normal(0, 5^2)$ priors. The same MCMC settings (warmup of 1000 iterations and 1000 saved iterations for each of four chains) were used.

For Bayesian double-dipping methods, the same training sample was used to select subgroups and to compute posterior subgroup intervals. The Bayesian methods were horserule BCF, Bayesian ridge regression, and an unpenalized linear model. The unpenalized linear approach used 1000 Dirichlet-weighted linear regressions as a Bayesian bootstrap approximation to the sampling distribution of the CATEs. For each Bayesian method, a depth-2 regression tree was fit to the posterior mean CATEs with \texttt{evtree} \citep{grubinger2014evtree} using 50 trees and 1000 iterations, using the MEPS covariates as predictors and the risk-neutral criterion $\lambda=1$. Samples of subgroup effects were obtained by averaging posterior CATE draws within each selected leaf. Importantly, the Bayesian methods never see the $N_{\mathrm{test}}$ held-out evaluation samples at any stage. The unpenalized linear regression method performed inference with the Bayesian bootstrap \citep{rubin1981bayesian} based on 1000 independently drawn weight vectors $w = (w_1, \ldots, w_{N_{\text{train}}}) \sim \Dirichlet(1, \ldots, 1)$, which in the correctly-specified setting produces intervals asymptotically equivalent to those obtained under a flat prior.

For non-Bayesian methods, random forest and lasso outcome regressions were used to estimate $\mu_1(x)$ and $\mu_0(x)$. For the random forest, we use the \texttt{regression\_forest} function fit to $(X, A)$ to get direct predictions for $Y_i$ rather than the causal forest, with $\texttt{mtry} = P$. The same defaults used for the realized utility experiment are otherwise used. For the lasso, we use \texttt{cv.glmnet}, which performs 10-fold cross validation across an automatically generated set of lasso penalties. For honest inference, the subgroup tree was selected from the Bayesian horserule BCF training fit and the random forest or lasso nuisance functions were evaluated on the test sample. For double-dipping inference, the random forest or lasso nuisance fits were used to select a depth-2 tree on the training sample itself and then to estimate subgroup effects on that same sample (like the Bayesian methods, the double-dipping non-Bayesian methods never see the evaluation samples).

The random forest and lasso subgroup effects were estimated using the robust randomized trial estimator
\[
  \widehat \tau(G)
  =
  \frac{1}{N_G}
  \sum_{i:X_i \in G}
  \left[
    \widehat \mu_1(X_i)-\widehat \mu_0(X_i)
    +
    \frac{\{A_i-e\}\{Y_i-\widehat \mu_{A_i}(X_i)\}}{e(1-e)}
  \right],
\]
with the known randomization probability $e=0.2$. The corresponding standard-error estimate was computed using the method of \citet{wager2016high} and converted into nominal 95\% normal intervals.

\paragraph{Metrics}

For every selected leaf, we computed the true subgroup effect by averaging the true CATEs of individuals in that leaf. We then recorded whether the nominal 95\% interval covered this true subgroup effect and the width of the interval. Figure~\ref{fig:post-selection} summarizes the resulting interval widths and empirical coverage rates across leaves and simulation replications.

A summary of the workflow for this experiment is given in Algorithm~\ref{alg:post-selection-workflow}.

\begin{algorithm}[t]
\caption{Post-selection and double-dipping simulation workflow}
\label{alg:post-selection-workflow}
\begin{algorithmic}[1]
\Require Processed MEPS data, linear and nonlinear fitted truths, replication count $B=200$, inflation factor $c$
\ForAll{truths $m \in \{\text{linear},\text{nonlinear}\}$}
    \For{$b = 1,\ldots,B$}
      \State Generate a training sample with $N_{\mathrm{train}}=1000$ and $A_i \sim \Bernoulli(0.2)$.
      \State Generate an independent inference sample with $N_{\mathrm{test}}=500$ for honest methods.
      \State Fit horserule BCF, Bayesian ridge, and linear regression to training sample.
      \ForAll{Bayesian double-dipping fits}
        \State Select a depth-2 subgroup tree from posterior mean CATEs.
        \State Compute effects within each selected leaf and form 95\% credible intervals.
      \EndFor
      \State Fit random forest and lasso outcome regressions to training sample.
      \State For honest variants, perform inference with held-out data on horserule BCF tree.
      \State For double-dipping variants, construct subgroups and estimate their effects on the training sample.
      \State Record interval width and coverage of the true subgroup effect for every subgroup.
    \EndFor
  \EndFor
\end{algorithmic}
\end{algorithm}

\section{Detecting Genuine Subgroup Structure}

We now perform a simulation experiment to assess whether the BRAIDS utility (i) is capable of discovering ``true'' subgroup effects when treatment effect heterogeneity has a genuine subgroup structure and (ii) leads to accurate uncertainty quantification for the discovered subgroup effects.

\paragraph{Data generation and replication}

Each replication contains $N = 500$ independent observations with $P = 10$ predictors. We generate $X_i \sim \Normal(0,\Sigma)$, where $\Sigma_{jk}=0.3^{|j-k|}$ except that $\Sigma_{12} = \Sigma_{21} = 0$. Treatment is randomized independently of the predictors as $A_i \sim \Bernoulli(0.5)$, and
\[
  Y_i=\mu(X_i)+A_i\tau(X_i)+\epsilon_i,
  \qquad \epsilon_i \sim \Normal(0,1).
\]
We use the prognostic function $\mu(x) = 0.5x_1 + 0.5x_3 + 0.25x_4^2 - 0.25x_5x_6 + 0.25\sin(x_7)$. The treatment effect is $\tau(x) = \tau_{k(x)}$, where $k(x) = 1 + 1(x_1>0) + 2\,1(x_2>0)$ and $(\tau_1,\tau_2,\tau_3,\tau_4) = M(-1.5,-0.5,0.5,1.5)$. This defines four equally probable leaves of a depth-2 tree and a population average treatment effect of zero. We use $M \in \{0.25, 0.30, 0.35, 0.40, 0.45, 0.50,0.75, 1.00, 1.25\}$, with 200 replications at each value.

\paragraph{Methods}
We compare Bayesian causal forests (BCF), Bayesian linear regression (BLR), causal forests (labelled CSF in the figures), and an $R$-learner using gradient boosted decision trees. BCF was fitted with \texttt{flexBART} with default priors. BLR set $\sigma_\mu,\sigma_\tau \sim \text{Exponential}(1)$ after standardizing outcomes, with flat priors on the other coefficients. Causal forests use \texttt{causal\_forest} with the known propensity $\widehat e_i = 0.5$ and package defaults, including 2000 trees, and \texttt{mtry = 10}. The $R$-learner uses \texttt{rboost} with the same known propensity and default tuning. The $R$-learner is included in recovery comparisons but omitted from uncertainty quantification because this implementation does not supply subgroup confidence intervals.

\paragraph{Subgroup trees}
For BCF and BLR, we apply the greedy CART version of BRAIDS to the posterior treatment-effect draws at $\lambda\in\{0,1,1.5,2\}$. For causal forests and the $R$-learner, only point estimates are used, so the comparison is restricted to the risk-neutral case $\lambda = 1$. All reported subgroup trees therefore have depth at most 2.

\paragraph{Metrics}
We evaluate recovery of the true subgroup structure using three metrics. The \emph{adjusted Rand index} ($\ARI$) \citep{vinh2009information} compares the true and estimated partitions in terms of which pairs of observations are assigned to the same subgroup, correcting for agreement expected by chance; $\ARI = 1$ when the subgroups are recovered exactly, while $\ARI = 0$ when the discovered subgroups agree with what one would get by chance. The pairwise partition error is the proportion of observation pairs that are placed together by exactly one of the true and estimated partitions; hence, smaller values are better and zero corresponds to exact recovery. Finally, we report the treatment-effect $R^2$,
\begin{math}
  R^2_\tau
  = \frac{\sum_{k, i : X_i \in G_k} \{\tau(G_k) - \tau(\sX)\}^2}{\sum_{i=1}^{N} \{\tau(X_i) - \tau(\sX)\}^2}.
\end{math}
This measures the proportion of the true treatment-effect heterogeneity explained by the estimated partition. In contrast to the adjusted Rand index, it does not penalize differences between the true and estimated partitions that preserve the treatment-effect-relevant structure. Values closer to one are better.
Here, $\tau(G_k)$ and $\tau(\sX)$ are the empirical averages of
the true treatment effects in selected leaf $G_k$ and in the full
simulated sample, respectively.

\begin{figure}[t]
  \includegraphics[width=1\textwidth]{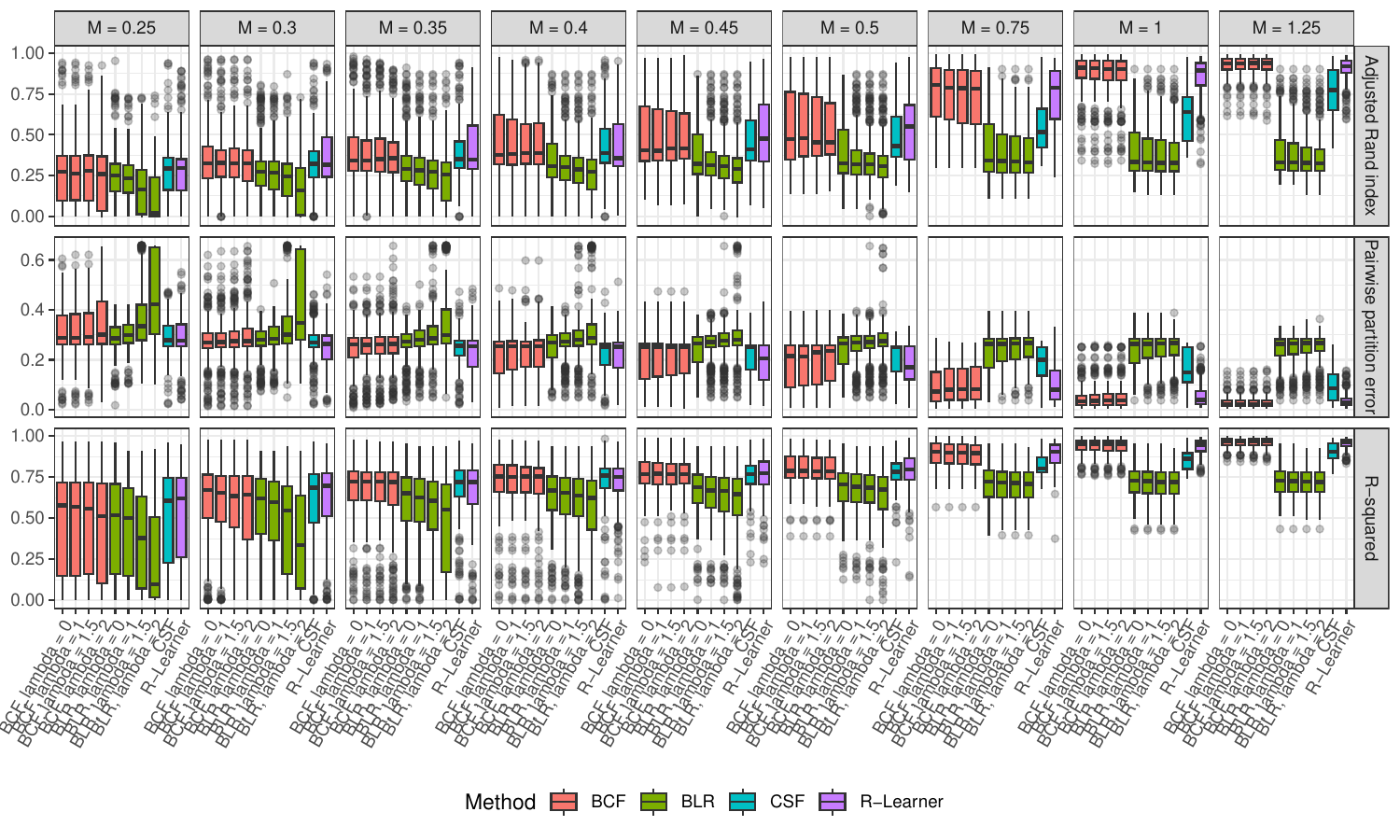}
  \caption{Results for recovering true subgroups with $N=500$ and $P=10$.
  Boxplots summarize 200 replications per effect size $M$ using the adjusted
  Rand index, pairwise partition error, and treatment-effect $R^2$.
  BCF and BLR use $\lambda\in\{0,1,1.5,2\}$; causal forests (CSF) and
  the $R$-learner use $\lambda=1$.\label{fig:subgroup_simulation}}
\end{figure}

\begin{figure}[t]
  \includegraphics[width=1\textwidth]{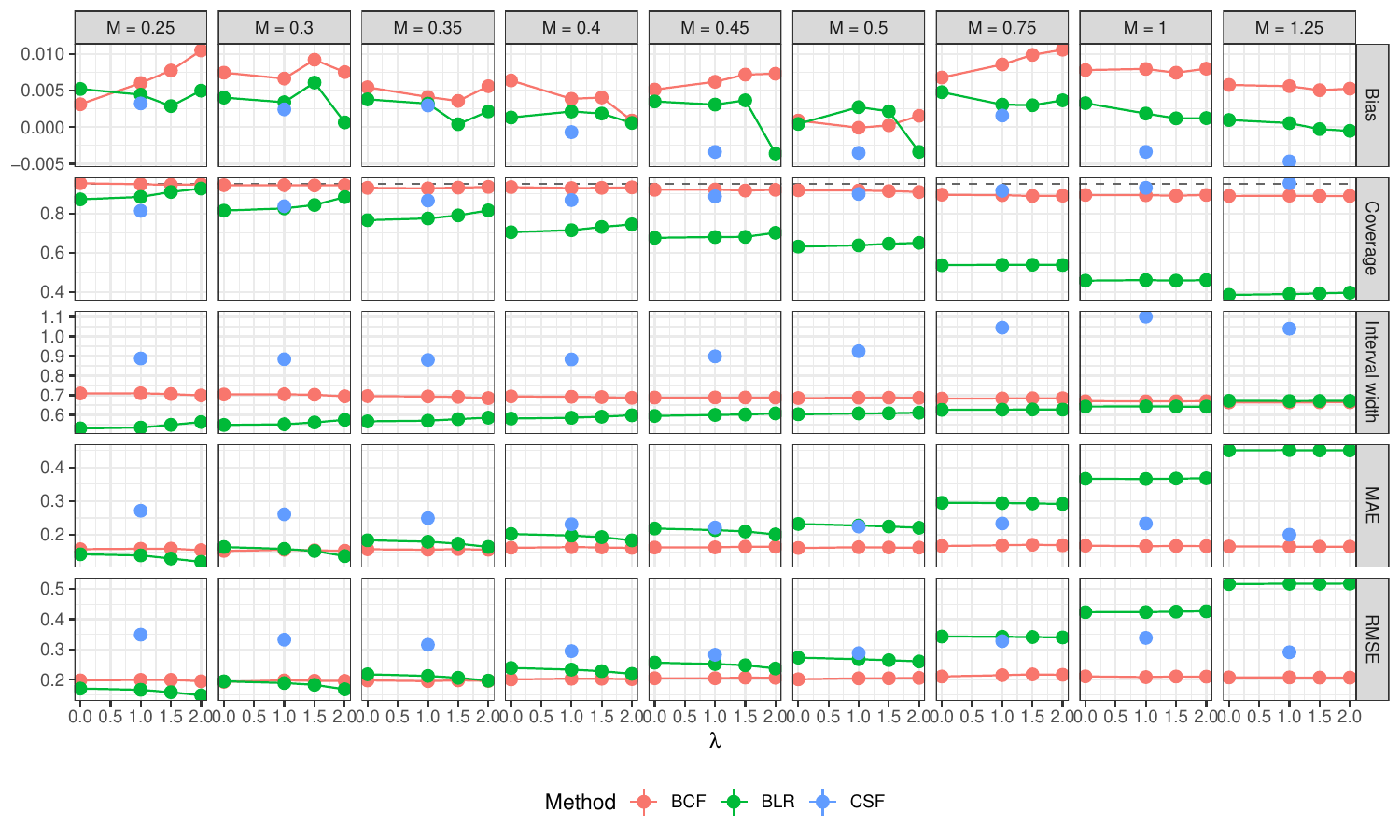}
  \caption{Estimation and uncertainty quantification in the selected subgroups
  with $N=500$ and $P=10$: bias, empirical 95\% interval coverage, mean interval
  width, MAE, and RMSE, pooled with equal weight per leaf over 200 replications
  per effect size $M$. BCF and BLR are evaluated at $\lambda\in\{0,1,1.5,2\}$;
  causal forests (CSF) appear only at $\lambda=1$. The dashed line marks 0.95
  coverage. The $R$-learner is omitted because subgroup confidence intervals
  are unavailable from this implementation.\label{fig:subgroup_simulation_uq}}
\end{figure}

\paragraph{Results}
Figure~\ref{fig:subgroup_simulation} displays the distributions across
replications of the three partition-recovery metrics for all four methods.
Figure~\ref{fig:subgroup_simulation_uq} reports bias, root mean squared error
(RMSE), mean absolute error (MAE), empirical coverage of nominal 95\% intervals,
and mean interval width for BCF, BLR, and causal forests. These quantities
pool the selected leaves across replications, giving each leaf equal weight;
coverage is the proportion of these intervals containing their leaf-specific
true effects. Both figures cover all nine values of $M$ and all stated
method-specific $\lambda$ values; the dashed coverage line marks 0.95.

Figure~\ref{fig:subgroup_simulation} shows that the BCF is surprisingly competitive with the causal forest in accurately identifying nearly-optimal subgroups, especially as the effect size grows; for large effects, the R-learner and BCFs perform roughly the same, and outperform both causal forests and Bayesian linear regression. Bayesian linear regression performs poorly relative to the tree based methods, suggesting (unsurprisingly) that detecting true subgroup structure is easier when the underlying algorithm is based on partitioning.

Figure~\ref{fig:subgroup_simulation_uq} shows that both the BCF and Bayesian linear regression tend to produce more accurate inferences than causal forests in the regimes where $M$ is small, which corresponds to the setting of most substantive interest for us. Additionally, both of these procedures produce narrower confidence intervals; in the case of Bayesian linear regression, this leads to undercoverage, especially for large values of $M$, while the BCF attains near-nominal coverage across most settings. Causal forests appear to perform the best when the effect sizes are large, reaching nominal coverage in those settings, but badly undercovering when the treatment effect sizes are small.

 We do not observe any obvious trends as $\lambda$ varies in Figure~\ref{fig:subgroup_simulation} for the BCF; it is much more important for Bayesian linear regression, which is misspecified, and in this case smaller values of $\lambda$ are associated with better performance on subgroup recovery. This trend also appears to hold somewhat for the BCF, but is less obvious and consistent. Figure~\ref{fig:subgroup_simulation_uq} suggests that, when $\lambda$ \emph{does} play a role for subgroup estimation accuracy, the tradeoff is poorer uncertainty quantification: in the small-$M$ settings, Bayesian linear regression has much better coverage when $\lambda$ is large. This may be due to the fact that Bayesian linear regression is solving an ``easier'' problem when it estimates poor subgroups, as the effects in these subgroups are likely close to null.

 Among the methods considered, when there is true subgroup structure, BCFs strike the best balance of accuracy of subgroup recovery and estimation accuracy. They are competitive with the R-learner on subgroup recovery while also providing uncertainty quantification close to the nominal level, falling to 90\% only when treatment effect heterogeneity is large; this is sensible, because the BCF prior has an inductive bias towards less treatment effect heterogeneity.

\section{Hybrid Optimization}

For $\lambda = 1$, we construct the exact depth-3 tree using the approach described by \citet{sverdrup2020policytree}. For $\lambda \ne 1$, however, exact optimization again becomes computationally expensive. In the canagliflozin trial we instead use the \emph{hybrid} or ``one-step-look-ahead'' approach to approximating the optimal decision tree when $\lambda \ne 1$. We describe the algorithm below.

Let $\mathfrak T_d(G)$ denote the collection of admissible decision trees of depth at most $d$ within subgroup $G$, subject to the minimum-leaf-size and candidate-split restrictions. For $T\in\mathfrak T_d(G)$, let $\mathscr P_T(G)$ denote its terminal-node partition and define
\[
  R_{\lambda,G}(T)
  =
  \sum_{A\in\mathscr P_T(G)}
  \sum_{i:X_i\in A}
  \left[
    \{\widehat\tau(A)-\widehat\tau(X_i)\}^2
    +(\lambda-1)\Var\{\tau(A)\mid\Data\}
  \right]
\]
where terms that are constant across candidate trees are omitted. The hybrid
depth-3 tree is constructed as follows:
\begin{enumerate}
\item Find an exact depth-2 tree at the root:
  \[
    T^{(0)}
    \in
    \underset{T\in\mathfrak T_2(\sX)}{\arg\min}\;
    R_{\lambda,\sX}(T).
  \]
\item If the root of $T^{(0)}$ is terminal, return the corresponding
  one-leaf tree. Otherwise, retain only its root splitting rule and discard
  the remaining splits of $T^{(0)}$. Let this rule partition $\sX$ into the
  two child nodes $G_L$ and $G_R$.
  
\item For each $b\in\{L,R\}$, fit an exact depth-2 subtree using only the
  observations in $G_b$:
  \[
    T^{(b)}
    \in
    \underset{T\in\mathfrak T_2(G_b)}{\arg\min}\;
    R_{\lambda,G_b}(T).
  \]
  If the optimizer selects a terminal node, leave $G_b$ unsplit.
  
\item Attach $T^{(L)}$ and $T^{(R)}$ to the retained root split.
\end{enumerate} 

In brief, CART is applied to choose splits, but instead of optimizing the usual criterion we choose splits by finding the best possible depth-2 subtree with the current node as the root. It is straightforward to extend this to depth $d > 3$, and it has total computation time of order $O(2^d \, N^2 \, P^2)$ rather than $O(N^d \, P^d)$ for exact optimization.

\section{Computation Time}

In Table~\ref{tab:timing} we provide the computation time required to fit models with different values of $N$ with $P$ set to $6$ for a $K = 4$ decision tree. We generated synthetic samples of $\tau(X_i)$ from a synthetic posterior centered at the step function
\begin{align*}
  \tau(x) = 1.25 1(x_1 > 0) - 0.75 1(x_2 > 0) + 0.5 1(x_1 > 0) 1(x_3 > 0) + \epsilon_x
\end{align*}
where $\epsilon_x \sim \Normal(0, 0.5^2)$, generated $2000$ posterior samples for each $\tau(X_i)$ where $X_i \sim \Normal(0, \Identity)$, and then fit BRAIDS with $\lambda = 1$ and $\lambda = 0$. We either computed the exactly-optimal decision tree or the CART approximation using the \texttt{braidstree} package included in the Supplementary Material.

\begin{table}
  \centering
  \begin{center}
    \singlespacing
    \begin{tabular}{lllrr}
      \toprule
      $N$  & $\lambda$ & Exact? & Time (Seconds) & Time (Relative) \\
      \midrule
      300  & 1         & No     & 0.01           & 1               \\
      300  & 1         & Yes    & 1.97           & 197             \\
      300  & 0         & No     & 0.03           & 3               \\
      300  & 0         & Yes    & 29.84          & 2984            \\
      \midrule
      3000 & 1         & No     & 0.30           & 1               \\
      3000 & 1         & Yes    & 1549.91        & 5163.3          \\
      3000 & 0         & No     & 0.98           & 3.3             \\
      3000 & 0         & Yes    & 36042.93       & 120140.0        \\
      \bottomrule
    \end{tabular}
  \end{center}
  \caption{Timing results.}
  \label{tab:timing}
\end{table}

Overall, we find that the CART approximation is quite scalable, while the exact decision trees are much more computationally intensive, especially for larger sample sizes. Because of the difficulty of fitting even an exact depth 2 tree, we used the hybrid approach (which requires fitting three exact depth 2 trees rather than one exact depth 3 tree) to find $\lambda \ne 1$ subgroups for the canagliflozin trial; in at least one case, this happened to coincide with the exact decision tree for $\lambda = 1$, suggesting that the hybrid approach is capable of also finding exactly-optimal decision trees.

We note that there is substantial room for improvement for the speed of our implementation of even the CART variant. For $N = 3000$, our CART implementation runs in 0.3 seconds, which is 30 times slower than \texttt{rpart} running the exact same algorithm; there is likely even more room for improvement for the exact methods. That being said, already the non-exact variants are highly scalable and can be run on trials much larger than the canagliflozin trial.

\section{Sensitivity Analysis for Canagliflozin Trial}

In this section, we show how different design choices for the canagliflozin trial impact the results. We consider varying $\lambda$ while keeping the Bayesian model (the Bayesian causal forest) fixed, as well as the impact of changing the underlying model while keeping $\lambda$ fixed.

\subsection{Sensitivity to the Choice of Estimation Method}

\begin{figure}[t]
  \centering
  \includegraphics[width=1\textwidth]{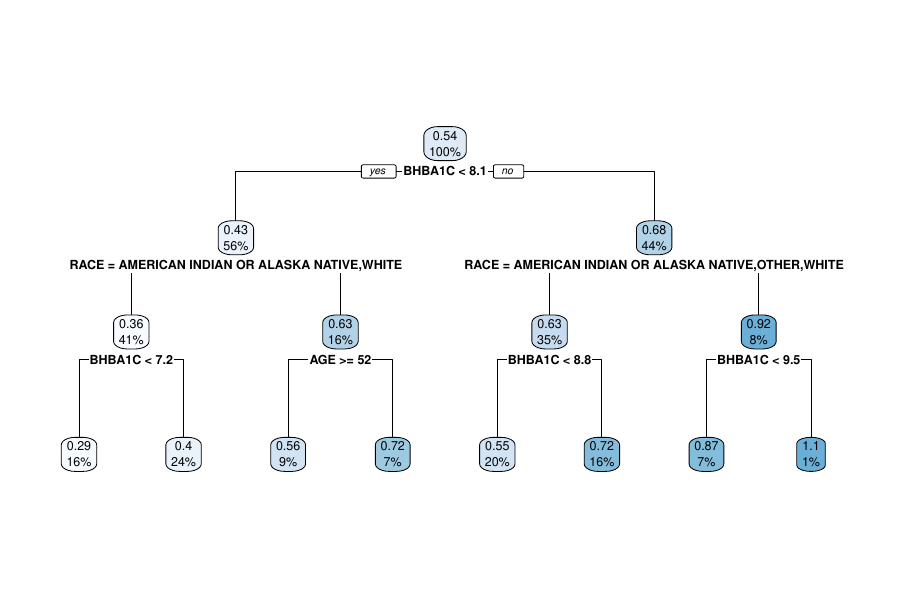}
  \caption{BRAIDS depth-3 decision tree under $\lambda = 1$ using the Bayesian ridge regression model for the canagliflozin trial.}
  \label{fig:ridge_1}
\end{figure}

We present the decision trees associated to BRAIDS with $\lambda = 1$ when one uses the Bayesian ridge regression model as the base model in Figure~\ref{fig:ridge_1}. The variables for the first two splits are the same as for the BCF model, except that HBA$_{1c}$ is used before race. There is, however, more divergence further down the tree, with age and HBA$_{1c}$ appearing in the bottom nodes rather than ethnicity and sex.

\begin{figure}[t]
  \centering
  \includegraphics[width=1\textwidth]{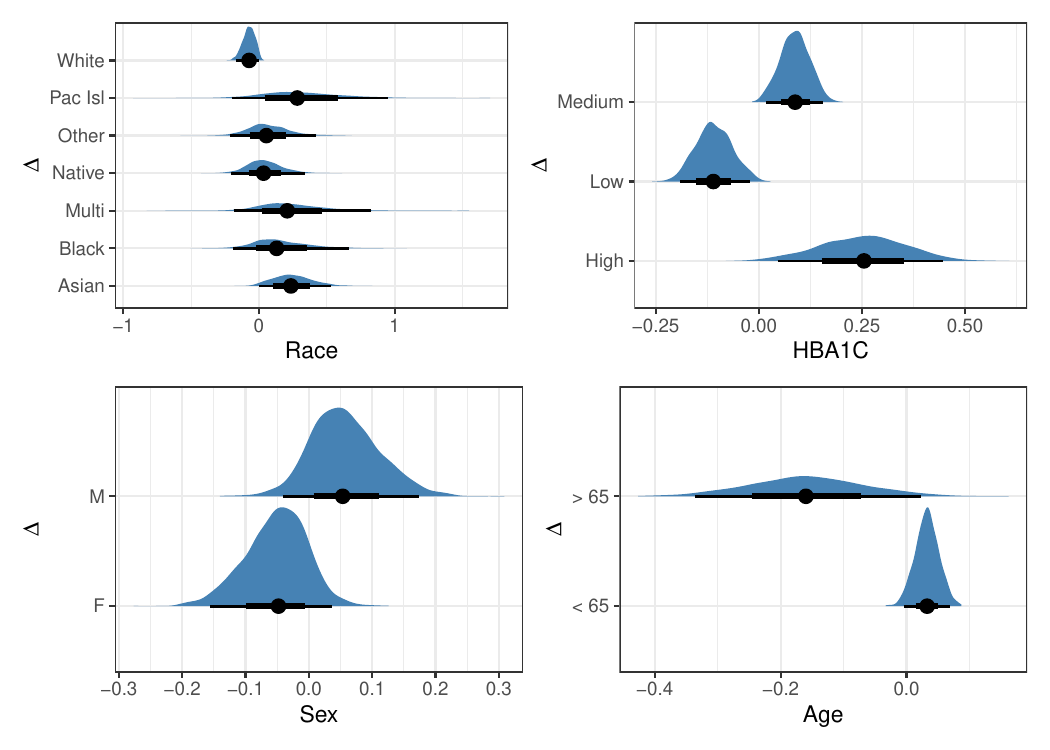}
  \caption{Posterior density for the deviation $\Delta$ from the overall average treatment effect within the prespecified subgroups defined by Race, HBA$_{1c}$, Sex, and Age.}
  \label{fig:ridge_2}
\end{figure}

\begin{figure}[t]
  \centering
  \includegraphics[width=1\textwidth]{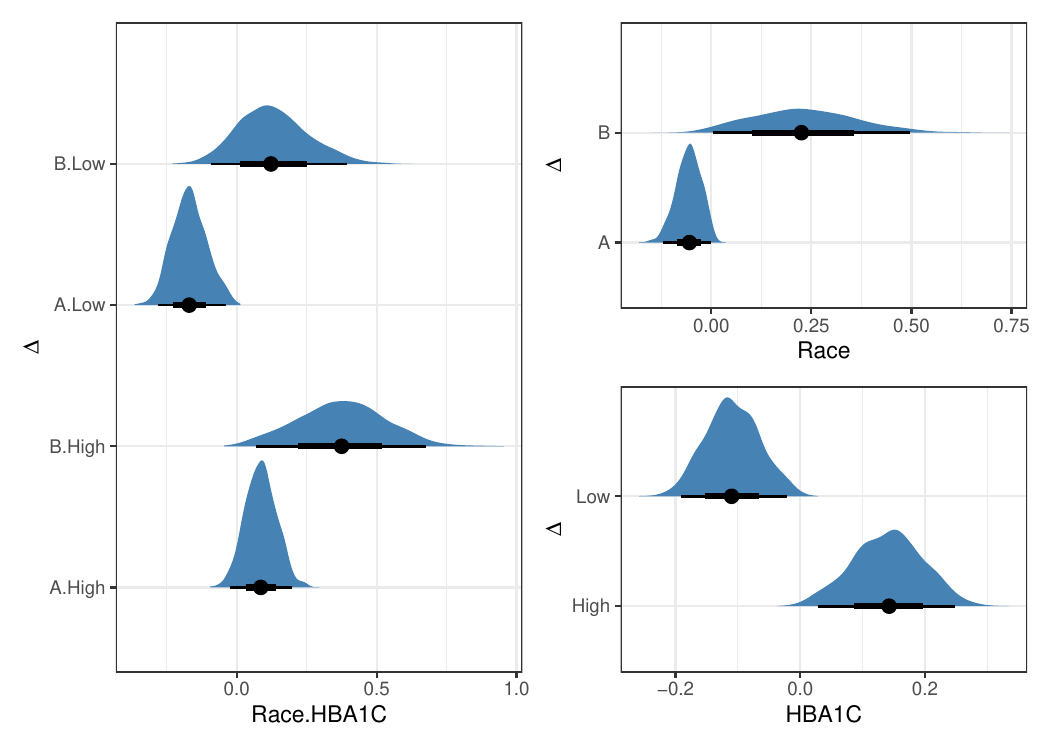}
  \caption{Posterior density for the deviation $\Delta$ from the overall average treatment effect within the subgroups identified by the BCF BRAIDS depth-3 decision tree under $\lambda = 1$.}
  \label{fig:ridge_3}
\end{figure}

Given this agreement, we proceed with a comparison of the same subgroups analyzed in the main manuscript, using the Bayesian ridge regression model in place of the BCF. These are given in Figure~\ref{fig:ridge_2} and Figure~\ref{fig:ridge_3}. Overall, we find that the results are not particularly sensitive to the choice of Bayesian ridge regression when compared with the analogous figures in the main manuscript.

\subsection{Sensitivity to the Utility}

We now examine how sensitive our results are to the choice of utility function and our choice of optimization algorithm. Each leaf node in our displayed trees shows the sample size in the node, the subgroup point estimate, and the subgroup standard deviation (except for $\lambda = 1$, as the standard deviation does not contribute to the utility).

For display purposes, we represent race and ethnicity with the following codes:
\begin{itemize}
\item \textbf{RACE:} American Indian or Alaskan Native (AI), Asian (AS), black (BL), Hawaiian (HA), multiple (MU), other (OT), and white (WH).
\item \textbf{ETHNIC:} Hispanic (H) or non-Hispanic (NH).
\end{itemize}

\begin{figure}
  \centering
  \includegraphics[width=1\textwidth]{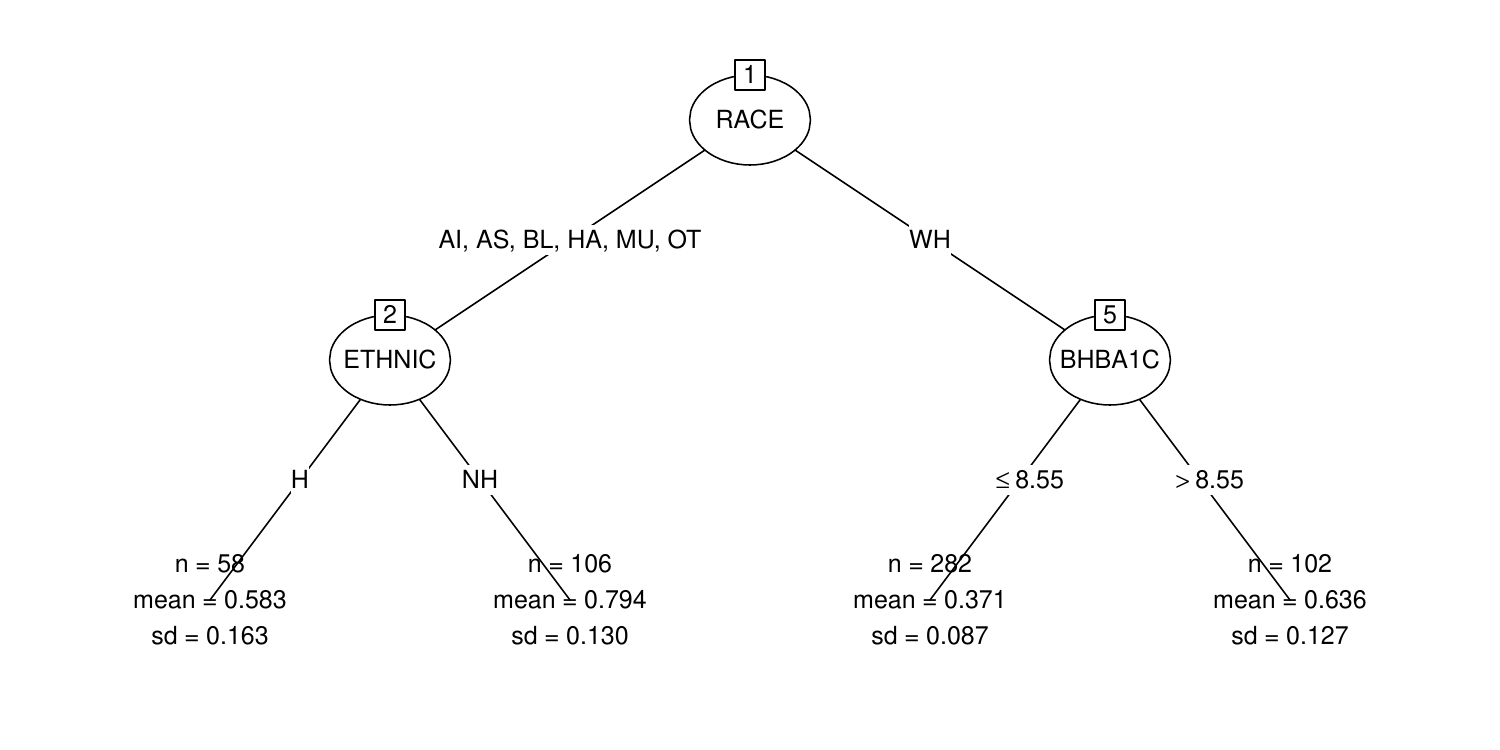}
  \caption{Exact depth-2 decision tree under $\lambda = 0$ for the canagliflozin trial.}
  \label{fig:exact_0}
\end{figure}

\begin{figure}
  \centering
  \includegraphics[width=1\textwidth]{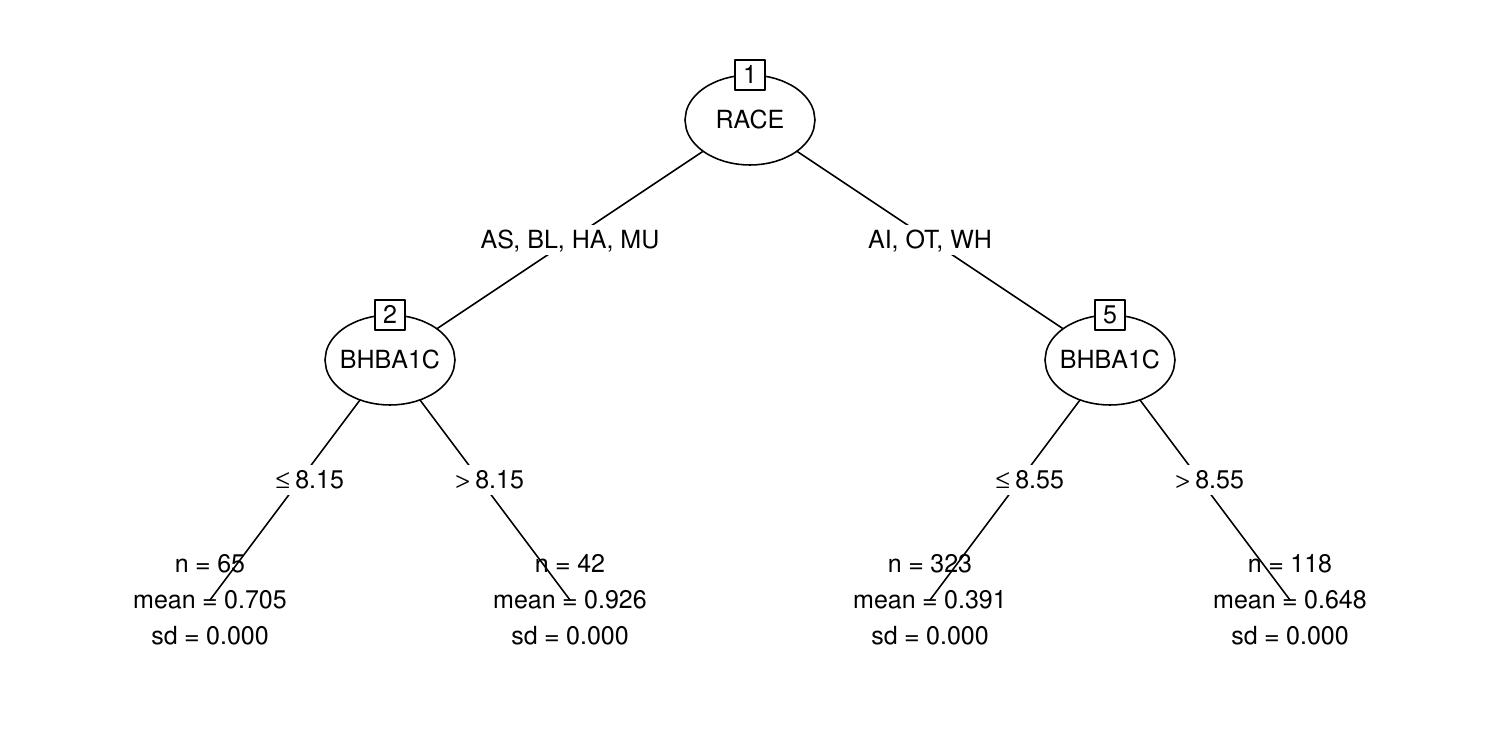}
  \caption{Exact depth-2 decision tree under $\lambda = 1$ for the canagliflozin trial.}
  \label{fig:exact_1}
\end{figure}

\begin{figure}
  \centering
  \includegraphics[width=1\textwidth]{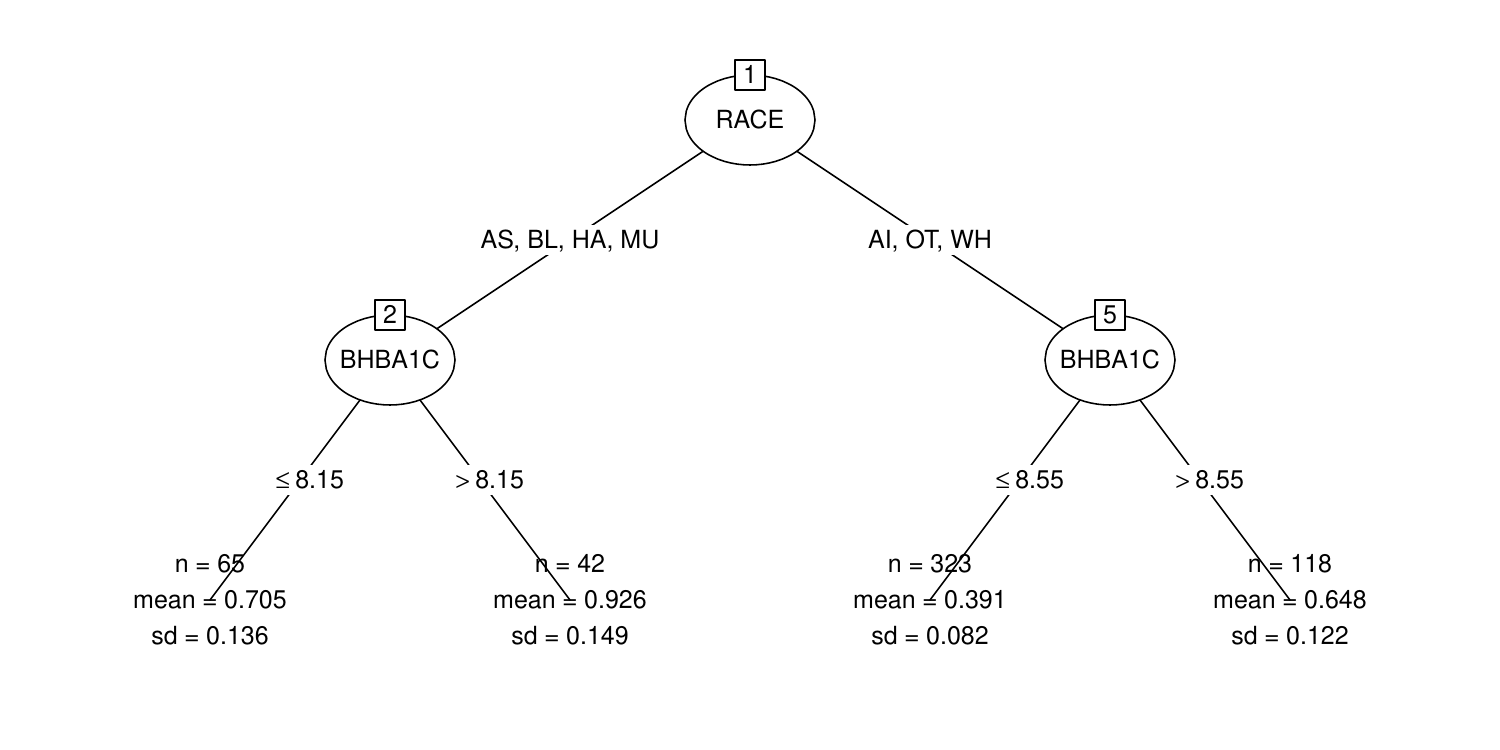}
  \caption{Exact depth-2 decision tree under $\lambda = 2$ for the canagliflozin trial.}
  \label{fig:exact_2}
\end{figure}

Figures~\ref{fig:exact_0}--\ref{fig:exact_2} show the exactly-optimal decision trees obtained from the Bayesian causal forest for depth $d = 2$; these are the same subgroups that are analyzed in Table~\ref{tab:ablation} of the main manuscript. The risk-seeking tree is unique in that not all subgroups are based on HBA$_{1c}$. The other two procedures find the same tree.

\begin{figure}
  \centering
  \includegraphics[width=1\textwidth]{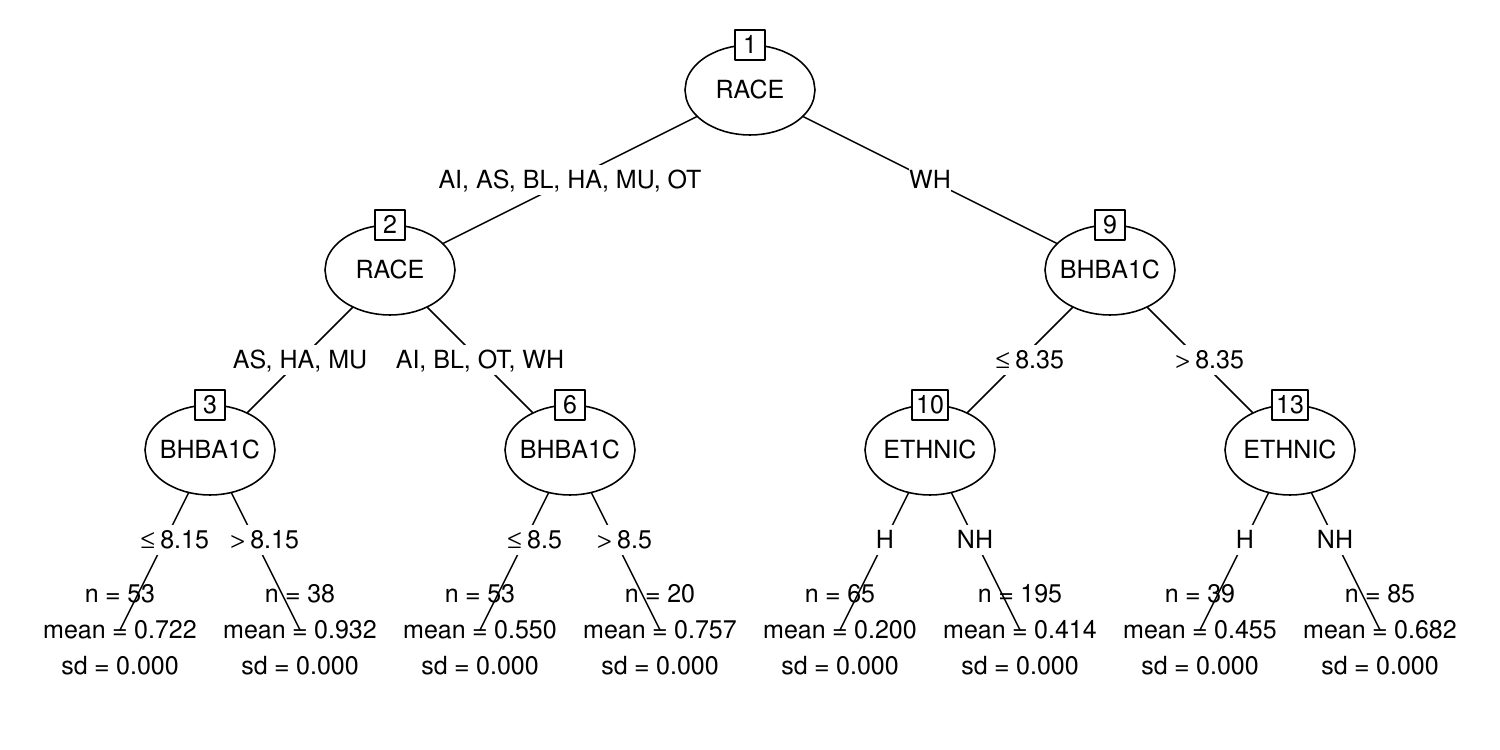}
  \caption{Exact depth-3 decision tree under $\lambda = 1$ for the canagliflozin trial.}
  \label{fig:exact_d3}
\end{figure}

Figure~\ref{fig:exact_d3} gives the only exact decision tree computed for depth $d = 3$, which is risk-neutral; this is also the exact same depth three tree found by the hybrid algorithm when $\lambda = 0$, although we note that this is not guaranteed to be optimal for this $\lambda$. All depth three exact or hybrid trees use race, HBA$_{1c}$, and ethnicity in their splits.

\begin{figure}
  \centering
  \includegraphics[width=1\textwidth]{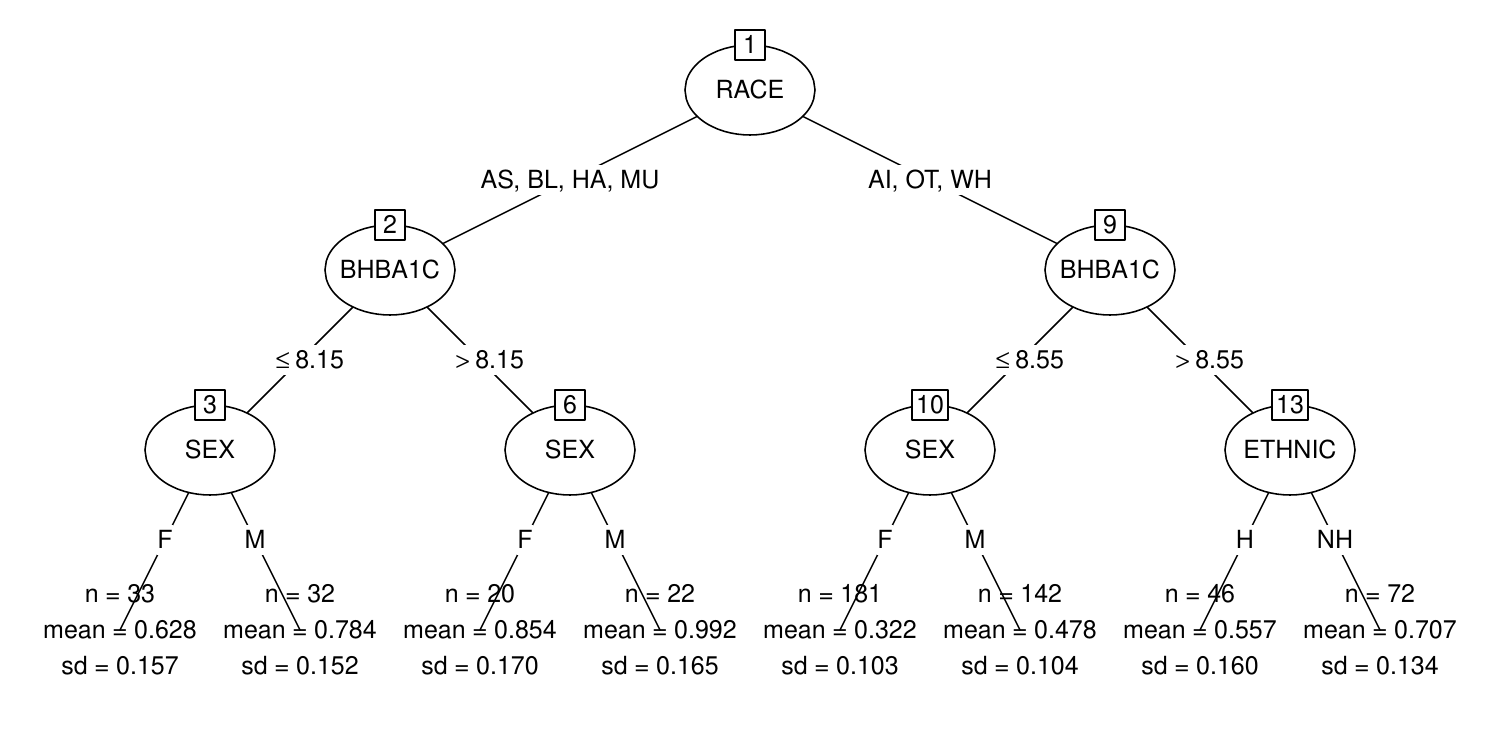}
  \caption{CART depth-3 decision tree under $\lambda = 0$ for the canagliflozin trial.}
  \label{fig:cart_0}
\end{figure}

\begin{figure}
  \centering
  \includegraphics[width=1\textwidth]{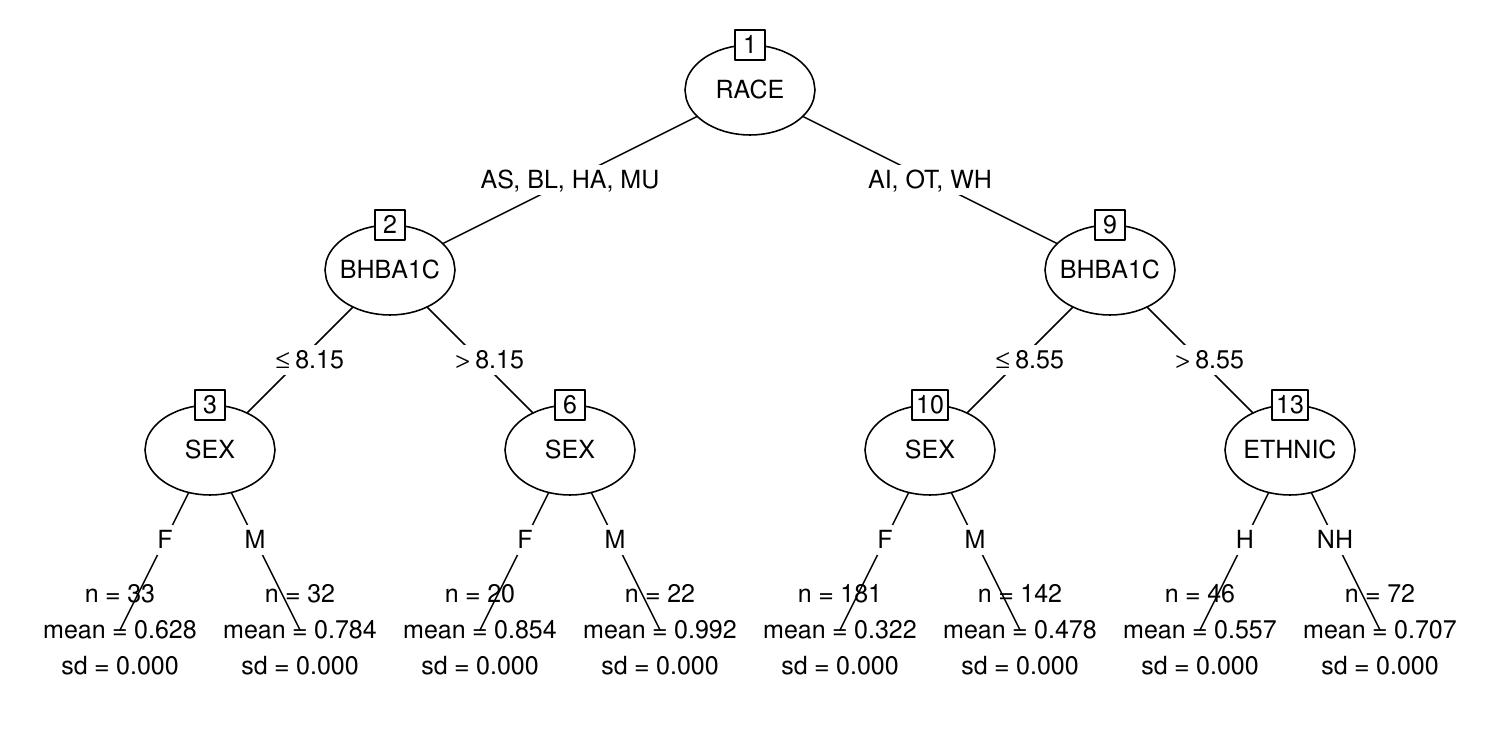}
  \caption{CART depth-3 decision tree under $\lambda = 1$ for the canagliflozin trial.}
  \label{fig:cart_1}
\end{figure}

\begin{figure}
  \centering
  \includegraphics[width=1\textwidth]{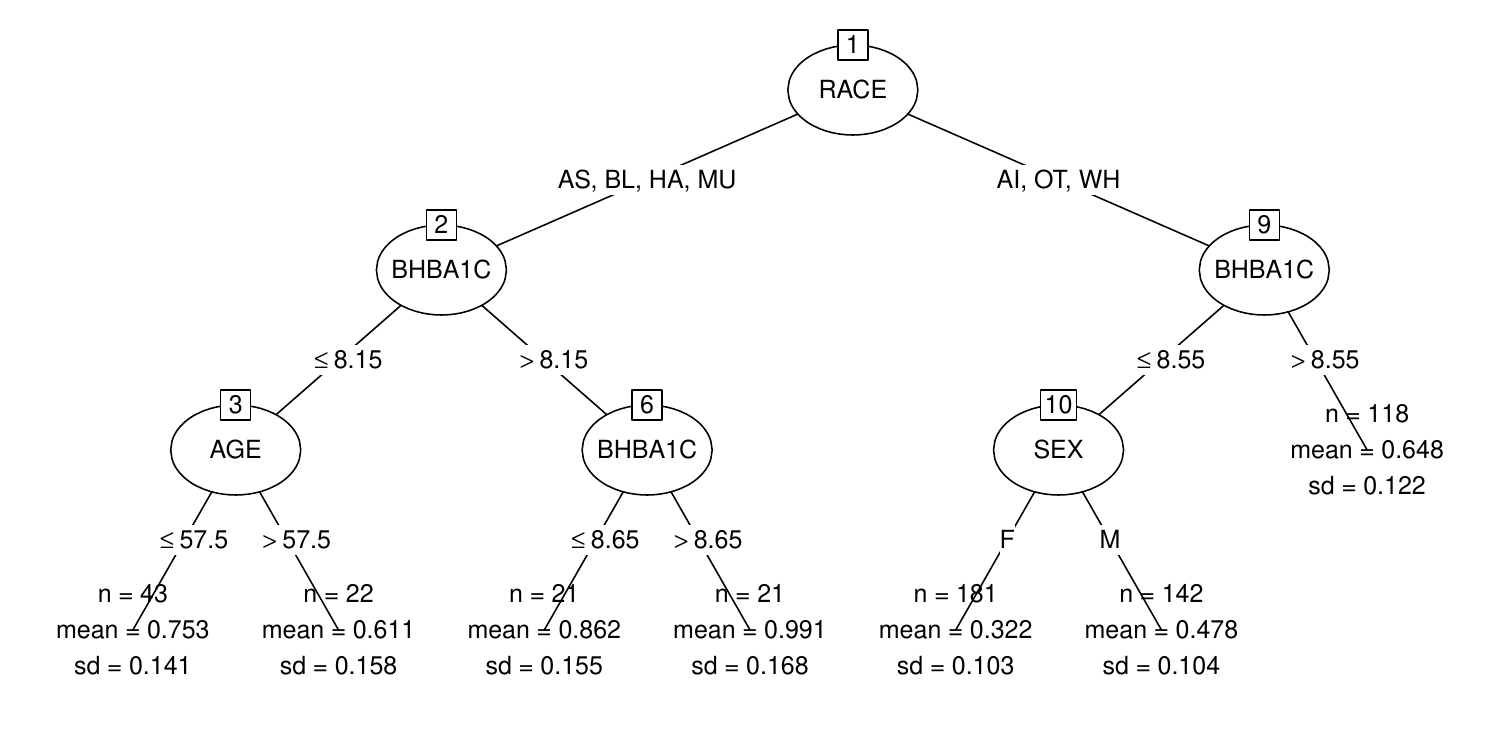}
  \caption{CART depth-3 decision tree under $\lambda = 2$ for the canagliflozin trial.}
  \label{fig:cart_2}
\end{figure}

Figures~\ref{fig:cart_0}--\ref{fig:cart_2} show the trees obtained using greedy optimization (CART) rather than the exact algorithm. The trees obtained here are generally different from the exact trees; they mostly use the same variables, but include sex more often, while $\lambda = 2$ uniquely includes age. Interestingly, we see that CART with $\lambda = 2$ does not grow a full tree; this is possible because when $\lambda > 1$ it is no longer the case that we are guaranteed to be able to increase the utility by performing more splits. 

\section{Aside on Policy Estimation}

Rather than searching for subgroups that merely \emph{differ} in their treatment effects, an alternative is to learn an \emph{individualized treatment rule} (ITR) that assigns treatment whenever the expected benefit outweighs its cost. Let $V: \mathcal X \to \{0, 1\}$ denote a policy that treats an individual with covariates $x$ when $V(x) = 1$. The empirical welfare of a policy can be written, up to an additive constant, as
\begin{align}
    \label{eq:empirical-welfare}
    U(V, \theta) = \sum_{i = 1}^N V(X_i) \{\tau(X_i) - \delta\},
\end{align}
where $\delta > 0$ encodes a per-unit treatment cost or minimum clinically important difference. Maximizing the posterior expected utility $R(V) = \mathbb{E}\{U(V, \theta) \mid \mathcal D\}$ is associated with the expected welfare maximization (EWM) principle of \citet{manski2004statistical};  \citet{kitagawa2018who} derive minimax optimal regret rate estimates of $V(x)$ when the propensity score is known, while \citet{athey2021policy} extend the approach to observational data using doubly‑robust scores. Alternatively, we might evaluate a procedure according to whether the treatment exceeds some efficacy threshold, without expressing a preference for how far the threshold is exceeded:
\begin{align}
    \label{eq:efficacy}
    U(V, \theta) = \sum_{i = 1}^N V(X_i) \, [1\{\tau(X_i) \ge \delta\} - c].
\end{align}
The value $c$ in \eqref{eq:efficacy} effectively corresponds to the local false positive rate we are willing to tolerate in determining whether the treatment is effective or not at $X_i$. 

Integrating \eqref{eq:empirical-welfare} and \eqref{eq:efficacy} with respect to the posterior distribution produces expected utilities $R(V) = \sum_{i = 1}^N V(X_i) \{\widehat \tau(X_i) - \delta\}$ and $R(V) = \sum_{i = 1}^N V(X_i) \, [\Pi\{\tau(X_i) \ge \delta \mid \mathcal D\} - c]$ respectively. As before, we can construct a penalized expected utility of the form $R(\vartheta) = R(V_\vartheta) + Q(\vartheta)$ where $\{V_\vartheta\}$ is a family of admissible policies (such as decision trees or rule lists) and $Q(\vartheta)$ penalizes the complexity of the policy (such as $Q(\mathcal T) = - \eta \, \operatorname{Depth}(\mathcal T)$ for decision trees).

\section{Cumulative-Sum Evaluation of BRAIDS}
\label{sec:cumulative-sum-proof}

In this section, we show how the utility function $R(G)$ associated to the BRAIDS utility can be evaluated efficiently for $\lambda = 1$. Specifically, we derive the split-gain identity and show how to use it to efficiently evaluate the utility. Let $N_B$ denote the number of observations in a nonempty node $B$ and let $S$ denote the number of joint posterior draws $\tau_i^{(s)}$ of $\tau(X_i)$ that we have. Throughout, we will identify a node $B$ with the set of $X_i$'s that are associated to that node, writing $X_i \in B$ to denote that $X_i$ is associated to node $B$ in the tree.

Define
\begin{align}
  \label{eq:supp-centered-draws}
  \widehat \tau_i = \frac{1}{S} \sum_{s = 1}^S \tau_i^{(s)}
  \quad \text{and} \quad
  r_i = \frac{1}{\sqrt{S - 1}}(\tau_i^{(1)} - \widehat \tau_i, \ldots, \tau_i^{(S)} - \widehat \tau_i)^\top
\end{align}
and define analogously for node $B$ the quantities
\begin{align*}
  \tau_B^{(s)}
  = \frac{1}{N_B}\sum_{i: X_i \in B}\tau_i^{(s)},\qquad
  \widehat\tau(B)
  = \frac{1}{N_B}\sum_{i: X_i\in B} \widehat \tau_i,\qquad
  \bar r_B
  = \frac{1}{N_B}\sum_{i: X_i \in B} r_i.
\end{align*}
Because $\widehat \tau(B) = S^{-1} \sum_s \tau_B^{(s)}$, we can approximate the posterior variance of $\tau(B)$ in terms of these quantities as
\begin{align}
  \label{eq:supp-node-variance}
  \widehat v_B
  = \frac{1}{S-1}\sum_{s=1}^S \{\tau_B^{(s)}-\widehat\tau(B)\}^2
  = \|\bar r_B\|_2^2.
\end{align}
Consequently, we can estimate the posterior variance $\Var\{\tau(B) \mid \Data\}$ using only these quantities, without needing to explicitly estimate the covariance matrix of the $\tau_i$'s.

To approximate the BRAIDS utility in \eqref{eq:two-step-heterogeneity-2}, we simply replace the posterior means and variances by the estimates above. Omitting the partition-independent constant, we have
\begin{align*}
  \label{eq:supp-mc-braids}
  \widehat R_\lambda(G) = \frac{1}{N} \sum_{k}
  \left[
  (1 - \lambda) \, N_{G_k} \|\bar r_{G_k}\|^2_2 -
  \sum_{i : X_i \in G_k} \{\widehat \tau_i - \widehat \tau(G_k)\}^2
  \right].
\end{align*}
Next, suppose that a node $A$ is split into two children $L$ and $R$ and let $\widehat \Delta_\lambda(L, R; A)$ be the change when $A$ is removed from $G$ and both $L$ and $R$ are added to $G$. Then
\begin{equation}
  \widehat\Delta_\lambda(L,R;A)
  = \frac{N_LN_R}{NN_A}
      \left[
        \{\widehat\tau(L)-\widehat\tau(R)\}^2
        +(1-\lambda)\|\bar r_L-\bar r_R\|_2^2
      \right].
  \label{eq:supp-split-gain}
\end{equation}
This follows from the fact that $\widehat R_\lambda(G)$ is in fact the utility function associated to a \emph{multivariate} regression tree under squared-error loss with the outcome being the expanded response $(\widehat \tau_i, \sqrt{1 - \lambda} r_i)^\top \in \Reals^{S + 1}$, with this change in utility being derived, for example, by \citet{ruan2006bi}.

\paragraph{Fast Evaluation of Candidate Splits}

We now show how to efficiently evaluate the gain of all potential splitting rules obtained by splitting a given node $A$ for a specified splitting variable; for simplicity, we will assume that the variable is continuous.
First, sort the $n = N_A$ observations in node $A$ as $i_1, \ldots, i_n$ by that variable. Define the scalar and vector cumulative sums
\begin{align*}
  a_m = \sum_{t=1}^m \widehat \tau_{i_t},
  \qquad
  b_m = \sum_{t=1}^m r_{i_t},
  \qquad m = 1, \ldots, n.
\end{align*}
For a split after observation $m$, the left and right means are obtained from $(a_m,b_m)$ and the totals $(a_n,b_n)$. In particular,
\begin{align}
  \label{eq:supp-cumulative-gain}
  \widehat \Delta_\lambda(m)
  &= \frac{m(n-m)}{Nn} \left[
    \left\{\frac{a_m}{m} -\frac{a_n-a_m}{n-m}\right\}^2
       + (1 - \lambda) \left\|\frac{b_m}{m} -\frac{b_n-b_m}{n-m}\right\|_2^2
     \right].
\end{align}
Moving to the next candidate threshold now only requires setting $a_{m+1} = a_m + \widehat\tau_{i_{m+1}}$ and $b_{m+1} = b_m + r_{i_{m+1}}$. This scheme can be directly applied with a ``greedy'' CART tree search, but is also useful for quickly evaluating all possible depth-$d$ decision trees by brute force.

\bibliographystyle{apalike}
\bibliography{references}